\documentclass[aps,twocolumn,showpacs]{revtex4}
\usepackage{graphicx}
\usepackage{dcolumn}
\usepackage{amsmath}
\usepackage{enumerate}

\begin{document}

\title{Uncertainties in nuclear transition matrix elements for neutrinoless 
$\beta\beta $ decay within the PHFB model}
\author{P. K. Rath$^{1}$, R. Chandra$^{1,2}$, K. Chaturvedi$^{3}$,
P. K. Raina$^{2} $ and J. G. Hirsch$^{4}$}

\affiliation{
$^{1}$Department of Physics, University of Lucknow, Lucknow-226007, India\\
$^{2}$Department of Physics and Meteorology, IIT, Kharagpur-721302, India\\
$^{3}$Department of Physics, Bundelkhand University, Jhansi-284128, India\\
$^{4}$Instituto de Ciencias Nucleares, Universidad Nacional Aut\'{o}noma de
M\'{e}xico, A.P. 70-543, M\'{e}xico 04510 D.F., M\'{e}xico}
\date{\today}

\begin{abstract}
The nuclear transition matrix elements $M^{\left( 0\nu \right) }$ for the 
neutrinoless double beta decay of $^{94,96}$Zr, $^{98,100}$Mo, 
$^{104}$Ru, $^{110}$Pd, $^{128,130}$Te and $^{150}$Nd isotopes 
in the case of $0^{+}\rightarrow 0^{+}$ transition 
are calculated using the PHFB wave functions, 
which are eigenvectors of 
four different parameterizations of a Hamiltonian with pairing plus multipolar
effective two-body interaction. 
Employing two (three) different parameterizations of Jastrow-type short
range correlations, 
a set of eight (twelve) different nuclear transition matrix
elements $M^{\left( 0\nu \right) }$ is built for each decay,
whose averages in conjunction with their standard deviations  
provide an estimate of the model uncertainties. 
\end{abstract}


\maketitle

\section{ INTRODUCTION}

Ascertaining the mass and nature of neutrinos 
requires the analysis of observational data obtained from three
complementary experiments, namely single-$\beta $ decay, neutrino
oscillation and neutrinoless double beta $\left( \beta \beta \right) _{0\nu
} $ decay. 
In any gauge theoretical model with 
spontaneous symmetry breaking, the observation of $\left( \beta \beta \right) _{0\nu }$
decay implies non-zero mass of Majorana neutrinos at the weak scale
independent of the underlying mechanisms \cite{sche82,hir97b}. The varied scope and 
far reaching nature of experimental as well as theoretical studies devoted to $\left(
\beta \beta \right) _{0\nu }$ decay over the
past decades have been 
excellently reviewed by Avignone \textit{et al.} \cite{avig08}
and references there in.

The observed
experimental limits on the half-life $T_{1/2}^{0\nu}$ of 
$\left( \beta \beta \right) _{0\nu }$ decay have already provided stringent 
limits on the associated gauge theoretical parameters \cite{klap06}. 
The reliability of extracted gauge theoretical parameters depends on the accuracy of 
nuclear transition matrix elements (NTMEs).
For a given transition, different NTMEs are obtained employing distinct nuclear models, 
and for a given model, they also depend on the model space and effective two-body 
interaction selected. Other uncertainties are related with the inclusion of
pseudoscalar and weak magnetism terms in the Fermi, Gamow-Teller
and tensorial NTMEs \cite{simk99,verg02}, 
finite size as well as short range correlations \cite{simk08,simk09,caur08,rath09},
and the use of two effective values of the axial-vector coupling constant $g_{A}$.

The spread between the calculated NTMEs provides a measure of the
theoretical uncertainty \cite{voge00}. In the case of the well studied $^{76}$Ge isotope, 
it was observed that the calculated decay rates differ by a factor of
6--7. The effective neutrino mass $\left\langle m_{\nu }\right\rangle $ is
inversely proportional to the square root of $T{_{1/2}^{0\nu }}$. Hence, the
uncertainty in the effective neutrino mass is about 2 to 3. For example, from
the experimental limit $T{_{1/2}^{0\nu }}$ $>1.6\times 10^{25}$ yr \cite{aals04},
the upper limits on $\left\langle m_{\nu }\right\rangle $ 
range between 0.4 eV and 1.0 eV, depending on the NTME \cite{haxt84,enge88,stau90}.
If the $\left( \beta \beta \right) _{0\nu }$ decay were observed in several nuclei,
the comparison of calculated ratios of the corresponding NTMEs-squared and the 
ratios of half-lives could also test the validity of nuclear structure calculations 
in a model independent way \cite{bile02}.

Rodin \textit{\textit{et al. }}\cite{rodi03} have estimated the theoretical
uncertainty employing two models, 
the QRPA and RQRPA, with three sets of basis states and three realistic two-body effective
interactions.
Different strategies to remove the sensitivity of QRPA calculations on the model parameters  
have been proposed  \cite{suho05,rodi06}. Further studies on uncertainties in
NTMEs due to short range correlations using the unitary correlation operator
method (UCOM) \cite{simk08} and self-consistent coupled cluster method
(CCM) \cite{simk09} have been also carried out by Faessler and coworkers.

Up to now, the QRPA model and its extensions have been the most
successful models in correlating the single-$\beta $ GT strengths and
half-lives of ($\beta ^{-}\beta ^{-}$)$_{2\nu }$ decay and the first in
explaining the observed suppression of $M_{2\nu }$ \cite{voge87,civi87}.
Nonetheless, the large scale shell model (LSSM)
calculations of Strasbourg-Madrid group are quite promising \cite{stma}. 
Deformation has been included at various levels of approximation in the QRPA formalism 
 \cite{pace04,yous09,alva04}. 
Recently, the effects
of pairing and quadrupolar correlations on the NTMEs of $\left( \beta
^{-}\beta ^{-}\right) _{0\nu }$ decay have also been studied in the
interacting shell model (ISM) \cite{caur08,caur08b} and the
projected-Hartree-Fock-Bogoliubov (PHFB)\ model \cite{chat08,chan09}.

The PHFB model, in conjunction with pairing plus quadrupole-quadrupole ($PQQ$) \cite
{bara68} interaction has been successful in the study of the $%
0^{+}\rightarrow 0^{+}$ transition of $\left( \beta ^{-}\beta ^{-}\right)
_{2\nu }$ decay, where it was possible to describe the lowest excited states
of the parent and daughter nuclei along with their electromagnetic
transition strengths, as well as to reproduce their measured $\beta \beta $
decay rates \cite{chan05,sing07}. The PHFB model is unique in allowing the description 
of the $\beta\beta$ decay in medium and heavy mass nuclei by projecting a set of states 
with good angular momentum, while treating the pairing and deformation degrees of freedom 
simultaneously and on equal footing.
On the other hand,  in the present version of the  PHFB model, the structure
of the intermediate odd $Z$-odd $N$ nuclei and hence, the single $\beta $
decay rates and the distribution of GT strength can not be studied. 
Notwithstanding this limitation,  
it is a convenient choice to examine the explicit role of
deformation on the NTMEs. In the study of $\beta ^{-}\beta ^{-}$ decay,
there are four noteworthy observations in connection with deformation
effects  \cite{chat08,chan09}, namely:
\begin{enumerate}[(i)]
\item There exists an inverse correlation between the
quadrupole deformation and the size of NTMEs $M_{2\nu }$, $M^{(0\nu)}$ and $%
M^{(0\nu)}_N$. 
\item The NTMEs are usually large in the absence of quadrupolar
correlations; they 
are almost constant for small admixture of the $QQ$ interaction and
substantially suppressed in deformed nuclei. 
\item In agreement with the observations made by \v{S}imkovic \textit{et al.} \cite{simk04}, 
the NTMEs have a well defined maximum when the deformation of parent and daughter 
nuclei are similar, and they are quite suppressed when the difference in the deformation is large.
\item The deformation effects are of equal importance in case of $\left( \beta
^{-}\beta ^{-}\right) _{2\nu }$ and $\left( \beta ^{-}\beta ^{-}\right)
_{0\nu }$ decay.
\end{enumerate} 

In earlier works, we have calculated NTMEs $M_{2\nu }$ for the $\left( \beta
^{-}\beta ^{-}\right) _{2\nu }$ \cite{chan05,sing07} and $M^{\left( 0\nu
\right) }$ for the $\left( \beta ^{-}\beta ^{-}\right) _{0\nu }$ decay \cite
{chat08} with the $PQQ$ effective interaction \cite{bara68}, and the effect of hexadecapolar 
correlations ($HH$) \cite{chan09} on the calculated spectroscopic properties and  
$\left( \beta ^{-}\beta ^{-}\right)$ decay rates has been studied.
In the present work, we employ two different parameterizations of the $QQ$ interaction, 
with and without the $HH$ correlations. Further, the NTMEs $M^{\left( 0\nu \right) }$ are 
calculated with three different parametrizations of Jastrow type of SRC employing 
the four sets of wave functions.
The twelve NTMEs provide a reasonable sample for estimating the associated
uncertainties.
In Sec II, the PHFB formalism employed to describe the $\left( \beta ^{-}\beta ^{-}\right) _{0\nu }$
decay with the inclusion of the finite size of the nucleons and short range correlations 
is shortly reviewed. In Sec. III, the four different parameterizations of the pairing plus 
multipole Hamiltonian are introduced, 
the calculated NTMEs vis-a-vis their radial evolution are analyzed, and their 
average values as well as standard deviations are estimated. Subsequently, the latter are employed to obtain 
upper limits on the effective mass of light Majorana neutrinos. Conclusions are given in Sec. IV.

\section{THEORETICAL FORMALISM}

In the Majorana neutrino mass mechanism, the inverse half-life of the $%
\left( \beta ^{-}\beta ^{-}\right) _{0\nu }$\ decay due to the exchange of
light neutrinos for the$\ 0^{+}\to 0^{+}$\ transition is given by \cite
{haxt84,doi85,tomo91} 
\begin{equation}
\left[T_{1/2}^{0\nu }(0^{+}\to 0^{+})\right]^{-1}=\left(\frac{\left\langle
m_{\nu }\right\rangle }{m_e}\right)^{2}G_{01}|M^{\left( 0\nu \right) }_{GT} -M^{\left( 0\nu \right) }_{F} |^{2},
\end{equation}
where the NTMEs $M^{\left( 0\nu \right) }_k$ are given by
\begin{equation}
M^{\left( 0\nu \right) }_k=\sum_{n,m}\left\langle 0_{F}^{+}\left\|
O_{k,nm}\tau _{n}^{+}\tau
_{m}^{+}\right\| 0_{I}^{+}\right\rangle ,
\end{equation}
with
\begin{equation}
O_{F} =  \left( \frac{g_{V}}{g_{A}} \right) ^{2}\, H(r_{12}),   
\,\,\,
O_{GT} =  
\mathbf{\sigma }_{1}\cdot \mathbf{\sigma }_{2}  \, H(r_{12})
\end{equation}
and
\begin{equation}
H(r_{12}) = \frac{R\phi (\overline{A}r_{12})}{r_{12}}.
\end{equation}
The origin of the neutrino potential $H(r_{12})$ is due to the exchange of light Majorana neutrinos
between nucleons being considered as point particles. 
To take the finite size of nucleons into account, neutrino potential $H(r_{12})$ is folded 
with a dipole form factor and rewritten as
\begin{equation}
H\left( r_{12}\right) =\frac{4\pi R}{\left( 2\pi \right) ^{3}}\int d^{3}q\frac{
\exp \left( i\mathbf{q}\cdot \mathbf{r_{12}}\right) }{q \left( q +
\overline{A}\right)} {\left( \frac{\Lambda ^{2}}{%
\Lambda ^{2}+q^{2}}\right) ^{4} },
\end{equation}
where 
\begin{equation}
\overline{A}=\left\langle E_{N}\right\rangle -\frac{1}{2}\left(
E_{I}+E_{F}\right) .
\end{equation}
and the cutoff momentum $\Lambda$= 850 MeV \cite{chat08}. 

The short range correlations (SRC) are produced by the repulsive nucleon-nucleon
potential generated through the exchange of $\rho $ and $\omega $ mesons.
They have been included in the calculations of $M^{\left (0\nu \right)}$ for the
$\left( \beta ^{-}\beta^{-}\right) _{0\nu }$
decay through the phenomenological Jastrow type of correlations with Miller-Spenser
parametrization \cite{mill76}, effective operators \cite{wu85}, exchange of
$\omega $-meson \cite{jghi95}, UCOM \cite{simk08,kort07}
and self-consistent CCM \cite{simk09}. It has been observed
that the effects due to the Jastrow type of correlations with Miller-Spenser
parametrization are usually strong \cite{wu85}, where as the UCOM and self-consistent
CCM have weak effects. Further, \v{S}imkovic \textit{et al.} \cite{simk09} have shown that
it is possible to parametrize the SRC effects of Argonne V18 and CD-Bonn two nucleon
potentials by the Jastrow type of correlations within a few percent accuracy.
Explicitly, the effects due to the
SRC can be incorporated in the calculation of $M^{\left (0\nu \right)}$ through
the prescription
\begin{equation}
O_k \rightarrow f O_k f   ,
\end{equation}
with
\begin{equation}
f(r)=1-ce^{-ar^{2}}(1-br^{2}),
\end{equation}
where $a=1.1$, $1.59$ and $1.52$ $fm^{-2}$, $b=0.68$, $1.45$ and $1.88$ $%
fm^{-2}$ and $c=1.0$, $0.92$ and $0.46$ for Miller-Spencer, Argonne V18 and
CD-Bonn NN potentials, respectively. In the next section, the NTMEs
$M^{\left( 0\nu \right) }$ are calculated in the PHFB
model by employing these three sets of parameters for the SRC,
denoted as SRC1, SRC2 and SRC3, respectively.

\begin{figure}[htbp]
\includegraphics [scale=0.7,angle=270]{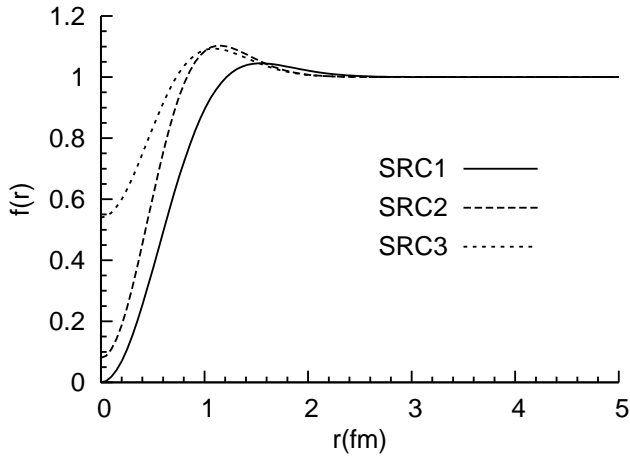} 
\caption{Radial dependence of $f(r)$ for the three different parameterizations of the SRC.}
\label{fig1}
\end{figure}

The three functions $f(r)$ are plotted in Fig. \ref{fig1}.
They have similar forms, but differ in its value at the origin, and at the position of 
its maximum, which lies at 1.54, 1.15 and 1.09 fm for SRC1, SRC2 and 
SRC3, respectively. 
They have clear influence on the radial evolution of the 
$\left(\beta^{-}\beta^{-}\right)_{0\nu}$ decay matrix elements discussed below.

The calculation of $M^{\left( 0\nu \right) }$ in the PHFB model has been
discussed in Ref. \cite{chat08} and one obtains the following expression for
NTMEs $M_{k }^{\left( 0\nu \right) }$ of $\left( \beta ^{-}\beta
^{-}\right) _{0\nu }$ decay 
\begin{eqnarray}
M_{k}^{\left( 0\nu \right) } &=&\left[ n^{Ji=0}n^{J_{f}=0}\right]
^{-1/2}  \nonumber \\
&&\times \int\limits_{0}^{\pi }n_{(Z,N),(Z+2,N-2)}(\theta
)\sum\limits_{\alpha \beta \gamma \delta }\left ( \alpha \beta \left|
O_{k}\right| \gamma \delta \right)   \nonumber \\
&&\times \sum\limits_{\varepsilon \eta }\frac{\left( f_{Z+2,N-2}^{(\pi
)*}\right) _{\varepsilon \beta }}{\left[ \left( 1+F_{Z,N}^{(\pi )}(\theta
)f_{Z+2,N-2}^{(\pi )*}\right) \right] _{\varepsilon \alpha }}  \nonumber
\\
&&\times \frac{\left( F_{Z,N}^{(\nu )*}\right) _{\eta \delta }}{\left[
\left( 1+F_{Z,N}^{(\nu )}(\theta )f_{Z+2,N-2}^{(\nu )*}\right) \right]
_{\gamma \eta }}sin\theta d\theta ,  \label{ntm}
\end{eqnarray}
where 
\begin{eqnarray}
n^{J} &=&\int\limits_{0}^{\pi }\left[ det\left( 1+F^{(\pi )}f^{(\pi
)^{\dagger }}\right) \right] ^{1/2}  \nonumber \\
&&\times \left[ det\left( 1+F^{(\nu )}f^{(\nu )^{\dagger }}\right) \right]
^{1/2}d_{00}^{J}(\theta )sin(\theta )d\theta ,
\end{eqnarray}
and 
\begin{eqnarray}
n_{(Z,N),(Z+2,N-2)}{(\theta )} &=&\left[ det\left( 1+F_{Z,N}^{(\nu
)}f_{Z+2,N-2}^{(\nu )^{\dagger }}\right) \right] ^{1/2}  \nonumber \\
&&\times \left[ det\left( 1+F_{Z,N}^{(\pi )}f_{Z+2,N-2}^{(\pi )^{\dagger
}}\right) \right] ^{1/2}.
\end{eqnarray}
The $\pi (\nu )$\ represents the proton (neutron) of nuclei involved in the $%
\left( \beta ^{-}\beta ^{-}\right) _{0\nu }$ decay process. The matrices $%
f_{Z,N}$ and $F_{Z,N}(\theta )$ are given by 
\begin{eqnarray}
f_{Z,N} &=&\sum\limits_{i}C_{ij_{\alpha },m_{\alpha }}C_{ij_{\beta
},m_{\beta }}\left( v_{im_{\alpha }}/u_{im_{\alpha }}\right) \delta
_{m_{\alpha },-m_{\beta }}, \\
F_{Z,N}(\theta ) &=&\sum\limits_{m_{\alpha }^{\prime }m_{\beta }^{\prime
}}d_{m_{\alpha },m_{\alpha }^{\prime }}^{j_{\alpha }}(\theta )d_{m_{\beta
},m_{\beta }^{\prime }}^{j_{\beta }}(\theta )f_{j_{\alpha }m_{\alpha
}^{\prime },j_{\beta }m_{\beta }^{\prime }}.
\end{eqnarray}
The extra factor 1/4 in the Eq. (28) of Ref. \cite{chat08} should not be there.

\section{RESULTS AND DISCUSSIONS}

The model space, single particle energies (SPE's), parameters of the $PQQ$ type
of effective two-body interaction and the method to fix them have been already given
in Refs. \cite{chan05,sing07,chat08}. Presently, we use the
effective Hamiltonian written as \cite{chan09} 
\begin{equation}
H=H_{sp}+V(P)+V(QQ)+V(HH),
\end{equation}
where $H_{sp}$, $V(P)$, $V(QQ)$ and $V(HH)$ denote the single particle
Hamiltonian, the pairing, quadrupole-quadrupole and
hexadecapole-hexadecapole parts of the effective two-body interaction,
respectively. 
The quadrupole-quadrupole part of the effective two-body interaction $V(QQ)$ has 
three terms, namely the proton-proton, the neutron-neutron 
and the proton-neutron ones, whose coefficients are denoted by $\chi _{2pp},\chi _{2nn}$ 
and $\chi _{2pn}$, respectively.
In Refs. \cite{chan05,sing07,chat08}, 
the strengths of the like particle components of the QQ interaction were 
taken as $ \chi _{2pp}=\chi _{2nn} = 0.0105$ MeV \textit{b}$^{-4}$, where \textit{b} 
is the oscillator parameter.
The strength of proton-neutron component of the $QQ$ interaction $\chi _{2pn}$
was varied so as to fit the experimental excitation energy of the $\ $2$^{+}$ state, 
\ $E_{2^{+}}$. In the present work, we also employ an alternative isoscalar parametrization 
of the quadrupole-quadrupole interaction, 
by taking $\chi _{2pp}=\chi _{2nn}=\chi _{2pn}/2$.  
In this case, the three parameters are varied together to fit $E_{2^{+}}$.
We will refer to these two parameterizations of the quadrupole-quadrupole interaction as 
$PQQ1$ and $PQQ2$.

Employing either method,  
the experimental excitation energies of $2^{+}$ state $E_{2^{+}}$ \cite{saka84} can be
reproduced within about 2\% accuracy. The maximum change in $E_{4^{+}}$ and $%
E_{6^{+}}$ energies with respect to $PQQ1$ interaction \cite{chan05,sing07}
is about 5\% and 18\%, respectively. 
The reduced $B(E2$:$0^{+}\to 2^{+})$ transition probabilities, deformation parameters 
$\beta _{2}$, static quadrupole moments $Q(2^{+})$ and gyromagnetic factors $g(2^{+})$
are in an overall agreement with the experimental data \cite{rama01,ragh89} for both 
the parametrizations. In the case of $PQQ2$ parametrization, the maximum change in the 
calculated NTMEs $M_{2\nu }$ for the $0^{+}\rightarrow 0^{+}$ transition with respect 
to $PQQ1$ parametrization is about 21\% but for $^{94}$Zr isotope.

\begin{table*}[htbp]
\caption{Calculated NTMEs $M^{\left( 0\nu \right) }$ in the PHFB model with
four different parameterizations of the effective two-body interaction, three
different parameterizations of SRC, 
with nucleons taken as point particles (P) or with a dipole form factor (F),
for the $\left( \beta^{-}\beta ^{-}\right) _{0\nu }$ \ decay of 
$^{94,96}$Zr, $^{98,100}$Mo, $^{104}$Ru, $^{110}$Pd, $^{128,130}$Te and $^{150}$Nd 
isotopes.}
\label{tab1}
\begin{tabular}{clccccccccccccccccc}
\hline\hline
Nuclei &  & \multicolumn{17}{c}{$M^{(0\nu) }$} \\ \cline{3-19}
\multicolumn{1}{l}{} &  & \multicolumn{3}{c}{P} &  & \multicolumn{5}{c}{P+S}
&  & F &  & \multicolumn{5}{c}{F+S} \\ \cline{3-5}\cline{7-11}\cline{15-19}
&  & $\overline{A}$ &  & $\overline{A}/2$ &  & SRC1 &  & SRC2 &  & SRC3 &  & 
&  & SRC1 &  & SRC2 &  & SRC3 \\ \hline
\multicolumn{1}{r}{$^{94}${\small Zr}} & $PQQ1$ & {\small 5.4382} &  & 
{\small 5.8729} &  & {\small 4.3021} &  & {\small 5.0644} &  & {\small 5.4097%
} &  & {\small 4.6891} &  & {\small 4.0690} &  & {\small 4.6639} &  & 
{\small 4.8383} \\ 
\multicolumn{1}{r}{} & $PQQHH1$ & {\small 5.0015} &  & {\small 5.3947} &  & 
{\small 3.9472} &  & {\small 4.6528} &  & {\small 4.9734} &  & {\small 4.3069%
} &  & {\small 3.7315} &  & {\small 4.2820} &  & {\small 4.4441} \\ 
\multicolumn{1}{r}{} & $PQQ2$ & {\small 5.1183} &  & {\small 5.5700} &  & 
{\small 4.1781} &  & {\small 4.8201} &  & {\small 5.1063} &  & {\small 4.4912%
} &  & {\small 3.9802} &  & {\small 4.4818} &  & {\small 4.6259} \\ 
\multicolumn{1}{r}{} & $PQQHH2$ & {\small 4.7626} &  & {\small 5.1356} &  & 
{\small 3.7492} &  & {\small 4.4266} &  & {\small 4.7348} &  & {\small 4.0955%
} &  & {\small 3.5424} &  & {\small 4.0708} &  & {\small 4.2266} \\ 
\multicolumn{1}{r}{} &  &  &  &  &  &  &  &  &  &  &  &  &  &  &  &  &  & 
\\ 
\multicolumn{1}{r}{$^{96}${\small Zr}} & $PQQ1$ & {\small 3.9517} &  & 
{\small 4.2741} &  & {\small 3.0829} &  & {\small 3.6622} &  & {\small 3.9257%
} &  & {\small 3.3828} &  & {\small 2.9068} &  & {\small 3.3590} &  & 
{\small 3.4923} \\ 
\multicolumn{1}{r}{} & $PQQHH1$ & {\small 3.9363} &  & {\small 4.2413} &  & 
{\small 3.0330} &  & {\small 3.6333} &  & {\small 3.9072} &  & {\small 3.3459%
} &  & {\small 2.8507} &  & {\small 3.3192} &  & {\small 3.4578} \\ 
\multicolumn{1}{r}{} & $PQQ2$ & {\small 3.7804} &  & {\small 4.0875} &  & 
{\small 2.9450} &  & {\small 3.5017} &  & {\small 3.7549} &  & {\small 3.2335%
} &  & {\small 2.7758} &  & {\small 3.2103} &  & {\small 3.3385} \\ 
\multicolumn{1}{r}{} & $PQQHH2$ & {\small 3.7035} &  & {\small 3.9891} &  & 
{\small 2.8470} &  & {\small 3.4158} &  & {\small 3.6753} &  & {\small 3.1442%
} &  & {\small 2.6745} &  & {\small 3.1182} &  & {\small 3.2497} \\ 
\multicolumn{1}{r}{} &  &  &  &  &  &  &  &  &  &  &  &  &  &  &  &  &  & 
\\ 
\multicolumn{1}{r}{$^{98}${\small Mo}} & $PQQ1$ & {\small 8.7743} &  & 
{\small 9.5345} &  & {\small 7.0859} &  & {\small 8.2350} &  & {\small 8.7484%
} &  & {\small 7.6507} &  & {\small 6.7322} &  & {\small 7.6297} &  & 
{\small 7.8884} \\ 
\multicolumn{1}{r}{} & $PQQHH1$ & {\small 8.1669} &  & {\small 8.8420} &  & 
{\small 6.5385} &  & {\small 7.6442} &  & {\small 8.1395} &  & {\small 7.0846%
} &  & {\small 6.1984} &  & {\small 7.0618} &  & {\small 7.3114} \\ 
& $PQQ2$ & {\small 8.8254} &  & {\small 9.5866} &  & {\small 7.1202} &  & 
{\small 8.2806} &  & {\small 8.7992} &  & {\small 7.6907} &  & {\small 6.7630%
} &  & {\small 7.6695} &  & {\small 7.9307} \\ 
& $PQQHH2$ & {\small 8.0911} &  & {\small 8.7589} &  & {\small 6.4723} &  & 
{\small 7.5712} &  & {\small 8.0636} &  & {\small 7.0154} &  & {\small 6.1344%
} &  & {\small 6.9925} &  & {\small 7.2406} \\ 
&  &  &  &  &  &  &  &  &  &  &  &  &  &  &  &  &  &  \\ 
$^{100}${\small Mo} & $PQQ1$ & {\small 8.5939} &  & {\small 9.2939} &  & 
{\small 6.8691} &  & {\small 8.0522} &  & {\small 8.5763} &  & {\small 7.4413%
} &  & {\small 6.5036} &  & {\small 7.4282} &  & {\small 7.6920} \\ 
& $PQQHH1$ & {\small 8.2130} &  & {\small 8.8577} &  & {\small 6.5174} &  & 
{\small 7.6763} &  & {\small 8.1915} &  & {\small 7.0822} &  & {\small 6.1597%
} &  & {\small 7.0654} &  & {\small 7.3248} \\ 
& $PQQ2$ & {\small 8.6571} &  & {\small 9.3633} &  & {\small 6.9212} &  & 
{\small 8.1116} &  & {\small 8.6391} &  & {\small 7.4972} &  & {\small 6.5534%
} &  & {\small 7.4838} &  & {\small 7.7493} \\ 
& $PQQHH2$ & {\small 7.4186} &  & {\small 7.9968} &  & {\small 5.8774} &  & 
{\small 6.9312} &  & {\small 7.3994} &  & {\small 6.3904} &  & {\small 5.5520%
} &  & {\small 6.3756} &  & {\small 6.6113} \\ 
&  &  &  &  &  &  &  &  &  &  &  &  &  &  &  &  &  &  \\ 
$^{104}${\small Ru} & $PQQ1$ & {\small 6.2757} &  & {\small 6.7734} &  & 
{\small 4.9743} &  & {\small 5.8753} &  & {\small 6.2705} &  & {\small 5.4007%
} &  & {\small 4.6942} &  & {\small 5.3989} &  & {\small 5.5975} \\ 
& $PQQHH1$ & {\small 5.7976} &  & {\small 6.2339} &  & {\small 4.5484} &  & 
{\small 5.4102} &  & {\small 5.7895} &  & {\small 4.9596} &  & {\small 4.2809%
} &  & {\small 4.9548} &  & {\small 5.1454} \\ 
& $PQQ2$ & {\small 5.9034} &  & {\small 6.3698} &  & {\small 4.6777} &  & 
{\small 5.5267} &  & {\small 5.8989} &  & {\small 5.0789} &  & {\small 4.4137%
} &  & {\small 5.0777} &  & {\small 5.2647} \\ 
& $PQQHH2$ & {\small 5.3786} &  & {\small 5.7803} &  & {\small 4.2143} &  & 
{\small 5.0176} &  & {\small 5.3711} &  & {\small 4.5974} &  & {\small 3.9648%
} &  & {\small 4.5931} &  & {\small 4.7708} \\ 
&  &  &  &  &  &  &  &  &  &  &  &  &  &  &  &  &  &  \\ 
$^{110}${\small Pd} & $PQQ1$ & {\small 10.1361} &  & {\small 11.0441} &  & 
{\small 8.1250} &  & {\small 9.5068} &  & {\small 10.1167} &  & {\small %
8.7918} &  & {\small 7.6982} &  & {\small 8.7783} &  & {\small 9.0850} \\ 
& $PQQHH1$ & {\small 8.5617} &  & {\small 9.2893} &  & {\small 6.7742} &  & 
{\small 7.9988} &  & {\small 8.5408} &  & {\small 7.3694} &  & {\small 6.3963%
} &  & {\small 7.3535} &  & {\small 7.6262} \\ 
& $PQQ2$ & {\small 9.7208} &  & {\small 10.5944} &  & {\small 7.7929} &  & 
{\small 9.1163} &  & {\small 9.7011} &  & {\small 8.4328} &  & {\small 7.3842%
} &  & {\small 8.4187} &  & {\small 8.7128} \\ 
& $PQQHH2$ & {\small 9.0246} &  & {\small 9.8138} &  & {\small 7.1864} &  & 
{\small 8.4447} &  & {\small 9.0023} &  & {\small 7.7985} &  & {\small 6.7982%
} &  & {\small 7.7816} &  & {\small 8.0621} \\ 
&  &  &  &  &  &  &  &  &  &  &  &  &  &  &  &  &  &  \\ 
$^{128}${\small Te} & $PQQ1$ & {\small 4.3415} &  & {\small 4.7394} &  & 
{\small 3.4372} &  & {\small 4.0474} &  & {\small 4.3219} &  & {\small 3.7417%
} &  & {\small 3.2499} &  & {\small 3.7258} &  & {\small 3.8639} \\ 
& $PQQHH1$ & {\small 4.8152} &  & {\small 5.2111} &  & {\small 3.7261} &  & 
{\small 4.4626} &  & {\small 4.7931} &  & {\small 4.0916} &  & {\small 3.4994%
} &  & {\small 4.0740} &  & {\small 4.2401} \\ 
& $PQQ2$ & {\small 5.1422} &  & {\small 5.6212} &  & {\small 4.1056} &  & 
{\small 4.8082} &  & {\small 5.1233} &  & {\small 4.4521} &  & {\small 3.8893%
} &  & {\small 4.4374} &  & {\small 4.5956} \\ 
& $PQQHH2$ & {\small 5.0701} &  & {\small 5.5058} &  & {\small 3.9637} &  & 
{\small 4.7118} &  & {\small 5.0477} &  & {\small 4.3351} &  & {\small 3.7336%
} &  & {\small 4.3172} &  & {\small 4.4860} \\ 
&  &  &  &  &  &  &  &  &  &  &  &  &  &  &  &  &  &  \\ 
$^{130}${\small Te} & $PQQ1$ & {\small 5.7440} &  & {\small 6.3018} &  & 
{\small 4.6613} &  & {\small 5.4025} &  & {\small 5.7319} &  & {\small 5.0177%
} &  & {\small 4.4319} &  & {\small 5.0103} &  & {\small 5.1753} \\ 
& $PQQHH1$ & {\small 4.9231} &  & {\small 5.3418} &  & {\small 3.8530} &  & 
{\small 4.5817} &  & {\small 4.9067} &  & {\small 4.2084} &  & {\small 3.6277%
} &  & {\small 4.1964} &  & {\small 4.3595} \\ 
& $PQQ2$ & {\small 5.6568} &  & {\small 6.2055} &  & {\small 4.5875} &  & 
{\small 5.3192} &  & {\small 5.6446} &  & {\small 4.9397} &  & {\small 4.3610%
} &  & {\small 4.9320} &  & {\small 5.0951} \\ 
& $PQQHH2$ & {\small 4.9115} &  & {\small 5.3304} &  & {\small 3.8459} &  & 
{\small 4.5714} &  & {\small 4.8951} &  & {\small 4.1999} &  & {\small 3.6218%
} &  & {\small 4.1879} &  & {\small 4.3503} \\ 
&  &  &  &  &  &  &  &  &  &  &  &  &  &  &  &  &  &  \\ 
$^{150}${\small Nd} & $PQQ1$ & {\small 4.1436} &  & {\small 4.5674} &  & 
{\small 3.3937} &  & {\small 3.9137} &  & {\small 4.1420} &  & {\small 3.6355%
} &  & {\small 3.2316} &  & {\small 3.6375} &  & {\small 3.7514} \\ 
& $PQQHH1$ & {\small 3.1506} &  & {\small 3.4603} &  & {\small 2.5501} &  & 
{\small 2.9650} &  & {\small 3.1478} &  & {\small 2.7448} &  & {\small 2.4208%
} &  & {\small 2.7447} &  & {\small 2.8359} \\ 
& $PQQ2$ & {\small 4.0499} &  & {\small 4.4632} &  & {\small 3.3160} &  & 
{\small 3.8249} &  & {\small 4.0483} &  & {\small 3.5526} &  & {\small 3.1574%
} &  & {\small 3.5546} &  & {\small 3.6661} \\ 
& $PQQHH2$ & {\small 3.2415} &  & {\small 3.5638} &  & {\small 2.6341} &  & 
{\small 3.0545} &  & {\small 3.2392} &  & {\small 2.8305} &  & {\small 2.5031%
} &  & {\small 2.8311} &  & {\small 2.9234} \\ \hline\hline
\end{tabular}
\end{table*}

\begin{table*}[htbp]
\caption{
Maximum and minimum relative change in the NTME $M^{\left( 0\nu \right) }$ (in \%), for 
all nuclei included in table I, due to the
use of a different average energy denominator ( second column), the inclusion of three different 
parameterizations of the SRC (SRC1, SRC2 and SRC3) with point nucleons (third to fifth column), 
the inclusion of finite size effect (F) (sixth column) 
and finite size effect plus SRC (F+SRC1, F+SRC2 and F+SRC3 in last three columns). 
In each row, the results employing one of the four
different parameterizations of the effective two-body interaction are displayed.
}
\label{tab2}
\begin{tabular}{lccccccccccccccc}
\hline\hline
Parametrizatios & $\overline{A}/2$ &  & \multicolumn{5}{c}{P+S} &  & F &  & 
\multicolumn{5}{c}{F+S} \\ \cline{4-8}\cline{12-16}
&  &~~  & SRC1 &~~  & SRC2 &~~  & SRC3 &~~  &  &~~  & SRC1 &~~  & SRC2 &~~  & SRC3 \\ 
\hline\hline
$PQQ1$ & \multicolumn{1}{l}{\small 7.9--10.2} &  & {\small 18.1--22.0} &  & 
{\small 5.5--7.3} &  & {\small 0.04--0.7} &  & {\small 12.3--13.9} &  & 
{\small 22.0--26.4} &  & {\small 12.2--15.0} &  & {\small 9.5--11.6} \\ 
$PQQHH1$ & \multicolumn{1}{l}{\small 7.5--9.8} &  & {\small 19.1--22.9} &  & 
{\small 5.9--7.7} &  & {\small 0.1--0.7} &  & {\small 12.9--15.0} &  & 
{\small 23.2--27.6} &  & {\small 12.9--15.7} &  & {\small 10.0--12.5} \\ 
$PQQ2$ & \multicolumn{1}{l}{\small 7.9--10.2} &  & {\small 18.1--22.1} &  & 
{\small 5.6--7.4} &  & {\small 0.04--0.7} &  & {\small 12.2--14.5} &  & 
{\small 22.0--26.6} &  & {\small 12.2--15.1} &  & {\small 9.5--11.7} \\ 
$PQQHH2$ & \multicolumn{1}{l}{\small 7.4--9.9} &  & {\small 18.7--23.1} &  & 
{\small 5.8--7.8} &  & {\small 0.1--0.8} &  & {\small 12.7--15.1} &  & 
{\small 22.8--27.8} &  & {\small 12.7--15.8} &  & {\small 9.8--12.2} \\ 
\hline\hline
\end{tabular}
\end{table*}

The $HH$ part of the effective interaction $V(HH)$ is given as \cite{chan09}
\begin{eqnarray}
V(HH)&=&-\left( \frac{\chi _{4}}{2}\right) \sum\limits_{\alpha \beta \gamma
\delta }\sum\limits_{\nu }(-1)^{\nu }\langle \alpha |q_{4\nu }|\gamma
\rangle \nonumber \\
&&\times \langle \beta |q_{4-\nu }|\delta \rangle \ a_{\alpha }^{\dagger
}a_{\beta }^{\dagger }\ a_{\delta }\ a_{\gamma },
\end{eqnarray}
with $q{_{4\nu }}=r^{4}Y_{4\nu }(\theta ,\phi )$. The relative
magnitudes of the parameters of the $HH$ part of the two body interaction
are calculated from a relation suggested by Bohr and Mottelson \cite{bohr98}%
. The approximate magnitude of these constants for isospin $T=0$ is given by 
\begin{equation}
\chi _{\lambda }=\frac{4\pi }{2\lambda +1}\frac{m\omega _{0}^{2}}{%
A\left\langle r^{2\lambda -2}\right\rangle }\,\,\,\,\,\,\,\,{for}\mathrm{{\,}%
\lambda =1,2,3,4\cdot \cdot \cdot }
\end{equation}
and the parameters for the $T=1$ case are approximately half of their $T=0$
counterparts. Presently, the value of $\chi _{4}=0.2442$ $\chi
_{2}A^{-2/3}b^{-4}$ for $T=1$, which is exactly half of the $T=0$ case.

We refer to the calculations which include the hexadecapolar term $HH$ as $PQQHH$. 
We end up with four different parameterizations
of the effective two-body interaction, namely $PQQ1$, $PQQHH1$, $PQQ2$ and 
$PQQHH2$.

\begin{table}[htbp]
\caption{Average NTMEs $\overline{M}^{(0\nu )}$ and uncertainties $\Delta
\overline{M}^{(0\nu )}$ for the $\left( \beta ^{-}\beta ^{-}\right) _{0\nu }$
decay of $^{94,96}$Zr, $^{98,100}$Mo, $^{110}$Pd, $^{128,130}$Te and $^{150}$%
Nd isotopes. Both bare and quenched values of $g_{A}$ are considered.}
\label{tab3}
\begin{tabular}{cccccccccc}
\hline\hline
$\beta ^{-}\beta ^{-}$ & $g_{A}$ &  & \multicolumn{3}{c}{Case I} &  &
\multicolumn{3}{c}{Case II} \\ \cline{4-6}\cline{8-10}
emitters &  &  & $\overline{M}^{(0\nu )}$ &  & $\Delta \overline{M}^{(0\nu )}
$ &  & $\overline{M}^{(0\nu )}$ &  & $\Delta \overline{M}^{(0\nu )}$ \\
\hline
$^{94}${\small Zr} & \multicolumn{1}{l}{\small 1.254} &  & {\small 4.2464} &
& {\small 0.3883} &  & {\small 4.4542} &  & {\small 0.2536} \\
& \multicolumn{1}{l}{\small 1.0} &  & {\small 4.6382} &  & {\small 0.4246} &
& {\small 4.8668} &  & {\small 0.2759} \\
&  &  &  &  &  &  &  &  &  \\
$^{96}${\small Zr} & \multicolumn{1}{l}{\small 1.254} &  & {\small 3.1461} &
& {\small 0.2778} &  & {\small 3.3181} &  & {\small 0.1243} \\
& \multicolumn{1}{l}{\small 1.0} &  & {\small 3.4481} &  & {\small 0.3085} &
& {\small 3.6376} &  & {\small 0.1424} \\
&  &  &  &  &  &  &  &  &  \\
$^{98}${\small Mo} & \multicolumn{1}{l}{\small 1.254} &  & {\small 7.1294} &
& {\small 0.6013} &  & {\small 7.4656} &  & {\small 0.3635} \\
& \multicolumn{1}{l}{\small 1.0} &  & {\small 7.8398} &  & {\small 0.6826} &
& {\small 8.2099} &  & {\small 0.4358} \\
&  &  &  &  &  &  &  &  &  \\
$^{100}${\small Mo} & \multicolumn{1}{l}{\small 1.254} &  & {\small 6.8749}
&  & {\small 0.6855} &  & {\small 7.2163} &  & {\small 0.4977} \\
& \multicolumn{1}{l}{\small 1.0} &  & {\small 7.5660} &  & {\small 0.7744} &
& {\small 7.9419} &  & {\small 0.5769} \\
&  &  &  &  &  &  &  &  &  \\
$^{110}${\small Pd} & \multicolumn{1}{l}{\small 1.254} &  & {\small 7.8413}
&  & {\small 0.8124} &  & {\small 8.2273} &  & {\small 0.6167} \\
& \multicolumn{1}{l}{\small 1.0} &  & {\small 8.6120} &  & {\small 0.9184} &
& {\small 9.0370} &  & {\small 0.7128} \\
&  &  &  &  &  &  &  &  &  \\
$^{128}${\small Te} & \multicolumn{1}{l}{\small 1.254} &  & {\small 4.0094}
&  & {\small 0.4194} &  & {\small 4.2175} &  & {\small 0.3074} \\
& \multicolumn{1}{l}{\small 1.0} &  & {\small 4.4281} &  & {\small 0.4601} &
& {\small 4.6571} &  & {\small 0.3355} \\
&  &  &  &  &  &  &  &  &  \\
$^{130}${\small Te} & \multicolumn{1}{l}{\small 1.254} &  & {\small 4.4458}
&  & {\small 0.5231} &  & {\small 4.6633} &  & {\small 0.4269} \\
& \multicolumn{1}{l}{\small 1.0} &  & {\small 4.9065} &  & {\small 0.5837} &
& {\small 5.1459} &  & {\small 0.4802} \\
&  &  &  &  &  &  &  &  &  \\
$^{150}${\small Nd} & \multicolumn{1}{l}{\small 1.254} &  & {\small 3.1048}
&  & {\small 0.4649} &  & {\small 3.2431} &  & {\small 0.4434} \\
& \multicolumn{1}{l}{\small 1.0} &  & {\small 3.4334} &  & {\small 0.5181} &
& {\small 3.5856} &  & {\small 0.4952} \\ \hline\hline
\end{tabular}
\end{table}

\begin{table*}[htbp]
\caption{Extracted limits on effective light Majorana neutrino mass $%
\left\langle m_{\nu }\right\rangle $ and predicted half lives using average
NTMEs $\overline{M}^{(0\nu )}$ and uncertainties $\Delta \overline{M}^{(0\nu
)}$ for the $\left( \beta ^{-}\beta ^{-}\right) _{0\nu }$ decay of $^{94,96}$%
Zr, $^{98,100}$Mo, $^{110}$Pd, $^{128,130}$Te and $^{150}$Nd isotopes.}
\label{tab4}
\begin{tabular}{ccccccccccccccc}
\hline\hline
$\beta ^{-}\beta ^{-}$ & $g_{A}$ &  & $\overline{M}^{(0\nu )}$ &  & {\small %
ISM} &  & {\small (R)QRPA} &  & {\small IBM} & $G_{01}$ & $T_{1/2}^{0\nu }(%
\text{ }yr)$ & {\small Ref.} & $\left\langle m_{\nu }\right\rangle $ & $%
T_{1/2}^{0\nu }($y$)$ \\
emitters &  &  &  &  & \cite{caur08} &  & \cite{simk09} &  & \cite{bare09} &
$(\text{{\small 10}}^{-14}${\small y}$^{-1})$ &  &  &  & $<m_{\nu
}>=50\,\,meV$ \\ \hline
$^{94}${\small Zr} & {\small 1.254} &  & {\small 4.45}$\pm ${\small 0.25} &
&  &  &  &  &  & {\small 0.\allowbreak 1\thinspace 684} & {\small 1.9}$%
\times ${\small 10}$^{19}$ & \cite{arno99} & \multicolumn{1}{l}{{\small 6.41}%
$_{-0.35}^{+0.39}\times 10^{2}$} & {\small 3.13}$_{-0.33}^{+0.39}\times $%
{\small 10}$^{27}$ \\
& {\small 1.0} &  & {\small 4.87}$\pm ${\small 0.28} &  &  &  &  &  &  &  &
&  & \multicolumn{1}{l}{{\small 9.23}$_{-0.49}^{+0.56}\times 10^{2}$} &
{\small 6.48}$_{-0.68}^{+0.80}\times ${\small 10}$^{27}$ \\
&  &  &  &  &  &  &  &  &  &  &  &  &  &  \\
$^{96}${\small Zr} & {\small 1.254} &  & {\small 3.32}$\pm ${\small 0.12} &
&  &  & {\small 1.43--2.12} &  &  & {\small 5.\thinspace 930} & {\small 1.0}$%
\times ${\small 10}$^{21}$ & \cite{arno99} & \multicolumn{1}{l}{{\small 20.00%
}$_{-0.72}^{+0.78}$} & {\small 1.60}$_{-0.11}^{+0.13}\times ${\small 10}$%
^{26}$ \\
& {\small 1.0} &  & {\small 3.64}$\pm ${\small 0.14} &  &  &  &  &  &  &  &
&  & \multicolumn{1}{l}{{\small 28.70}$_{-1.08}^{+1.17}$} & {\small 3.29}$%
_{-0.24}^{+0.27}\times ${\small 10}$^{26}$ \\
&  &  &  &  &  &  &  &  &  &  &  &  &  &  \\
$^{98}${\small Mo} & {\small 1.254} &  & {\small 7.47}$\pm ${\small 0.36} &
&  &  &  &  &  & {\small 0.0018} & {\small 1.0}$\times ${\small 10}$^{14}$ &
\cite{frem52} & \multicolumn{1}{l}{{\small 1.62}$_{-0.08}^{+0.08}\times
10^{6}$} & {\small 1.06}$_{-0.10}^{+0.11}\times ${\small 10}$^{29}$ \\
& {\small 1.0} &  & {\small 8.21}$\pm ${\small 0.44} &  &  &  &  &  &  &  &
&  & \multicolumn{1}{l}{{\small 2.32}$_{-0.12}^{+0.13}\times 10^{6}$} &
{\small 2.16}$_{-0.21}^{+0.25}\times ${\small 10}$^{29}$ \\
&  &  &  &  &  &  &  &  &  &  &  &  &  &  \\
$^{100}${\small Mo} & {\small 1.254} &  & {\small 7.22}$\pm ${\small 0.50} &
&  &  & {\small 2.91--5.56} &  & {\small 3.732} & {\small 4.\thinspace 640}
& {\small 4.6}$\times ${\small 10}$^{23}$ & \cite{arno05} &
\multicolumn{1}{l}{{\small 0.48}$_{-0.03}^{+0.04}$} & {\small 4.32}$%
_{-0.54}^{+0.66}\times ${\small 10}$^{25}$ \\
& {\small 1.0} &  & {\small 7.94}$\pm ${\small 0.58} &  &  &  &  &  &  &  &
&  & \multicolumn{1}{l}{{\small 0.69}$_{-0.05}^{+0.05}$} & {\small 8.83}$%
_{-1.15}^{+1.44}\times ${\small 10}$^{25}$ \\
&  &  &  &  &  &  &  &  &  &  &  &  &  &  \\
$^{110}${\small Pd} & {\small 1.254} &  & {\small 8.23}$\pm ${\small 0.62} &
&  &  &  &  &  & {\small \allowbreak 1.\thinspace 422} & {\small 6.0}$\times
${\small 10}$^{17}$ & \cite{wint52} & \multicolumn{1}{l}{{\small 6.72}$%
_{-0.47}^{+0.54}\times 10^{2}$} & {\small 1.09}$_{-0.15}^{+0.18}\times $%
{\small 10}$^{26}$ \\
& {\small 1.0} &  & {\small 9.04}$\pm ${\small 0.71} &  &  &  &  &  &  &  &
&  & \multicolumn{1}{l}{{\small 9.63}$_{-0.70}^{+0.82}\times 10^{2}$} &
{\small 2.22}$_{-0.31}^{+0.40}\times ${\small 10}$^{26}$ \\
&  &  &  &  &  &  &  &  &  &  &  &  &  &  \\
$^{128}${\small Te} & {\small 1.254} &  & {\small 4.22}$\pm ${\small 0.31} &
& {\small 2.26} &  & {\small 3.21--5.65} &  & {\small 4.517} & {\small 0.1849%
} & {\small 1.1}$\times ${\small 10}$^{23}$ & \cite{arna03} &
\multicolumn{1}{l}{{\small 8.50}$_{-0.58}^{+0.67}$} & {\small 3.18}$%
_{-0.42}^{+0.52}\times ${\small 10}$^{27}$ \\
& {\small 1.0} &  & {\small 4.66}$\pm ${\small 0.34} &  &  &  &  &  &  &  &
&  & \multicolumn{1}{l}{{\small 12.10}$_{-0.81}^{+0.94}$} & {\small 6.44}$%
_{-0.84}^{+1.04}\times ${\small 10}$^{27}$ \\
&  &  &  &  &  &  &  &  &  &  &  &  &  &  \\
$^{130}${\small Te} & {\small 1.254} &  & {\small 4.66}$\pm ${\small 0.43} &
& {\small 2.04} &  & {\small 2.92--5.04} &  & {\small 4.059} & {\small %
4.\thinspace 490} & {\small 3.0}$\times ${\small 10}$^{24}$ & \cite{arna08}
& \multicolumn{1}{l}{{\small 0.30}$_{-0.02}^{+0.03}$} & {\small 1.07}$%
_{-0.17}^{+0.23}\times ${\small 10}$^{26}$ \\
& {\small 1.0} &  & {\small 5.15}$\pm ${\small 0.48} &  &  &  &  &  &  &  &
&  & \multicolumn{1}{l}{{\small 0.42}$_{-0.04}^{+0.04}$} & {\small 2.17}$%
_{-0.35}^{+0.47}\times ${\small 10}$^{26}$ \\
&  &  &  &  &  &  &  &  &  &  &  &  &  &  \\
$^{150}${\small Nd} & {\small 1.254} &  & {\small 3.24}$\pm ${\small 0.44} &
&  &  &  &  & {\small 2.321} & {\small \allowbreak 21.16} & {\small 1.8}$%
\times ${\small 10}$^{22}$ & \cite{argy09} & \multicolumn{1}{l}{{\small 2.55}%
$_{-0.31}^{+0.40}$} & {\small 4.69}$_{-1.06}^{+1.60}\times ${\small 10}$^{25}
$ \\
& {\small 1.0} &  & {\small 3.59}$\pm ${\small 0.50} &  &  &  &  &  &  &  &
&  & \multicolumn{1}{l}{{\small 3.63}$_{-0.44}^{+0.58}$} & {\small 9.49}$%
_{-2.16}^{+3.29}\times ${\small 10}$^{25}$ \\ \hline\hline
\end{tabular}
\end{table*}

\subsection{SRC and radial evolutions of NTMEs}

In Table~\ref{tab1}, the
NTMEs $M^{\left(0\nu \right) }$ evaluated using the HFB wave functions in conjunction 
with $PQQ1$, $PQQHH1$, $PQQ2$, $PQQHH2$ interactions and three different parametrizations 
of the Jastrow type of SRC 
for the nuclei $^{94,96}$Zr, $^{98,100}$Mo, $^{104}$Ru, $^{110}$Pd, $^{128,130}$Te and 
$^{150}$Nd are displayed.
The average energy
denominator $\overline{A}$ has been taken as $\overline{A}=1.12A^{1/2}$ MeV
following Haxton's prescription \cite{haxt84}. The NTMEs are calculated in the 
the approximations of point nucleons (P - 2nd and 3rd columns), finite size 
of nucleons (F - 7th column), point nucleons  with SRC (P+S - 4th to 6th columns), 
and finite size plus SRC (F+S - last three columns). To obtain additional 
information on the stability of the estimations of NTMEs $M^{\left( 0\nu
\right) }$, they are also calculated for $\overline{A}/2$ in the
energy denominator in the case of point nucleons, given in the column 3.

We present the relative changes in NTMEs $M^{\left( 0\nu \right) }$ (in \%) 
due to the different approximations in Table ~\ref{tab2}.
In each row, i.e. for each set of wave functions, the reference NTMEs 
$M^{\left( 0\nu \right) }$ are those calculated for point nucleons without SRC, 
given in the second column of Table ~\ref{tab1}. 
It can be observed that the relative change in NTMEs $M^{\left( 0\nu \right) }$, 
when the energy denominator is taken as $\overline{A}/2$ instead of $\overline{A}$, is 
of the order of 10 \%. It confirms that the dependence of NTMEs on average 
excitation energy $\overline{A}$ is small for the $\left( \beta ^{-}\beta ^{-}\right)_{0\nu}$ 
decay and the validity of the closure approximation is quite satisfactory.

The variation in $M^{\left( 0\nu \right) }$ due to the different parameterizations of 
the Hamiltonians (presented in the different rows) lies between 20--25\%.
It is noticed in general but for $^{128}$Te isotope that the NTMEs evaluated for both 
parameterizations of the quadrupolar interaction 
are quite close. The inclusion of the hexadecapolar term tends to reduce them by 
amounts which strongly depend on the specific nuclei.

The inclusion of SRC in the approximation of point nucleons (P+S) induces an extra 
quenching in the NTMEs $M^{\left(0\nu \right) }$, which can be of the order of 
18--23\% for SRC1, to negligible for SRC3.
The dipole form factor (F) always reduces the NTMEs by 12--15\% in comparision to  
the point particle case. 
Adding SRC (F+S) can further reduce the transition matrix elements, for SRC1, or slightly enhance 
them, partially compensating the effect of the dipole form factor. It is interesting to note 
that the effect of F-SRC2 is almost negligible, i.e., nearly the same as F. 

\begin{figure*}[htbp]
\begin{tabular}{cc}
\includegraphics [scale=0.32]{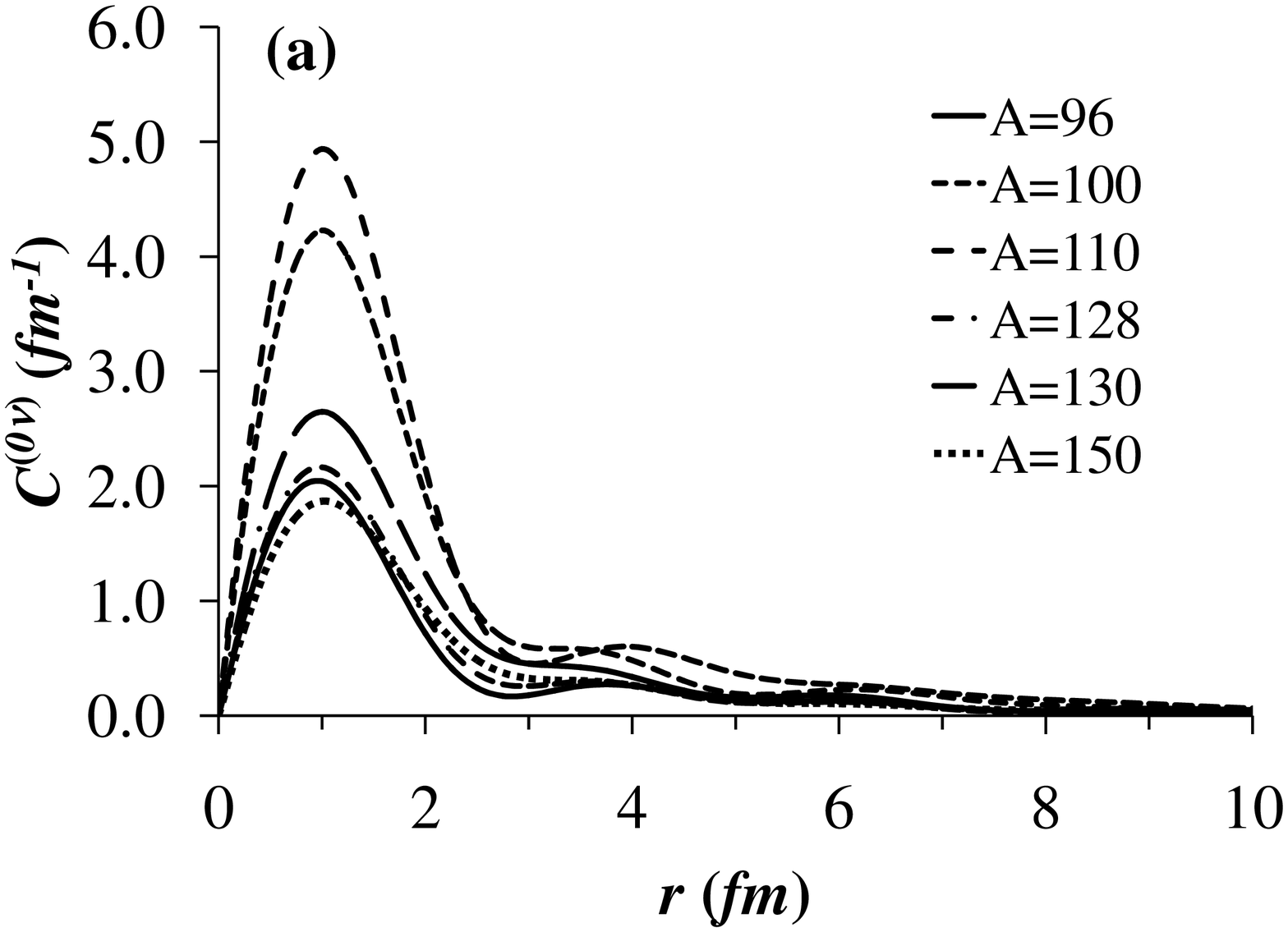} &
\includegraphics [scale=0.32]{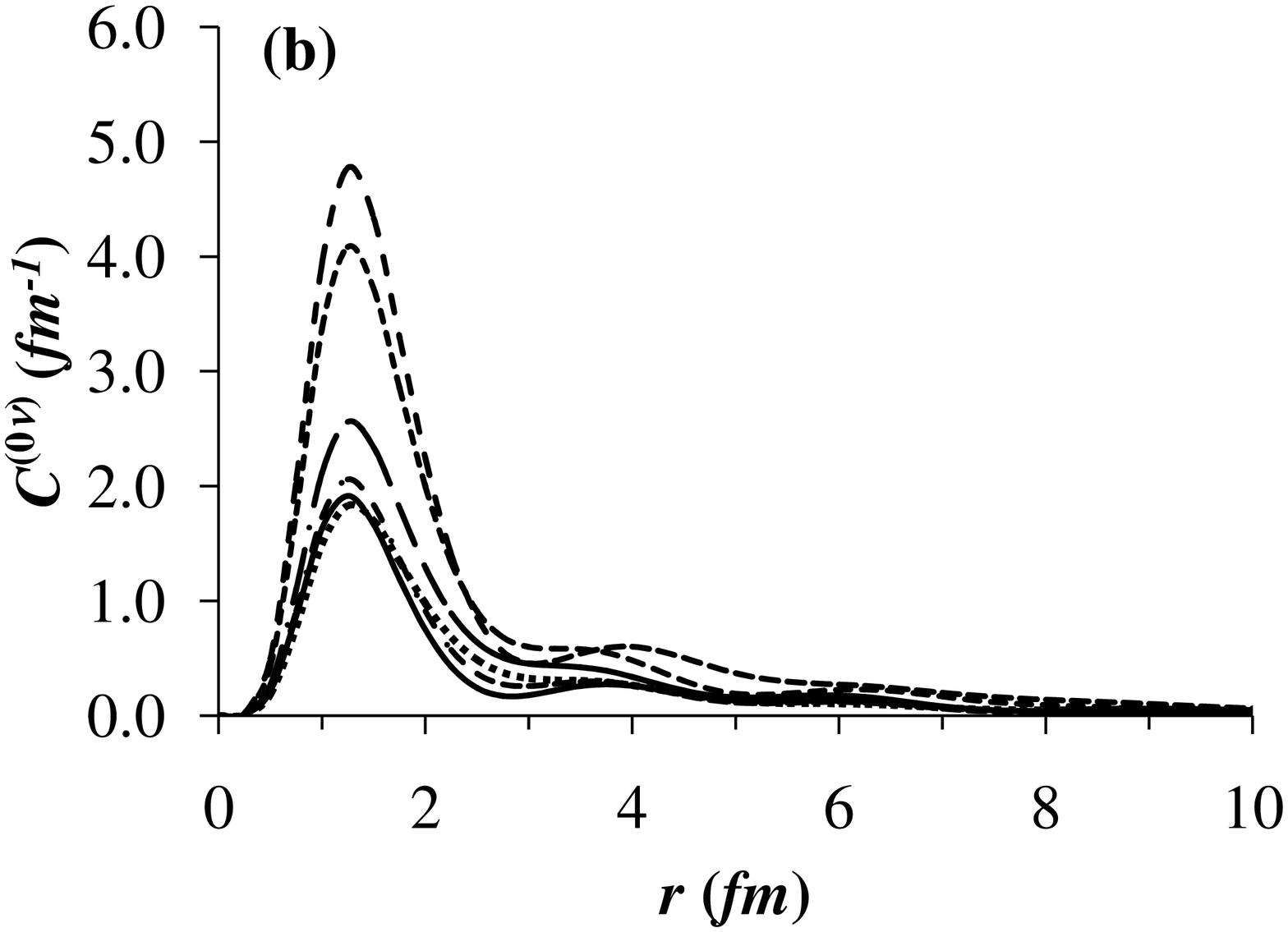} \\
\includegraphics [scale=0.32]{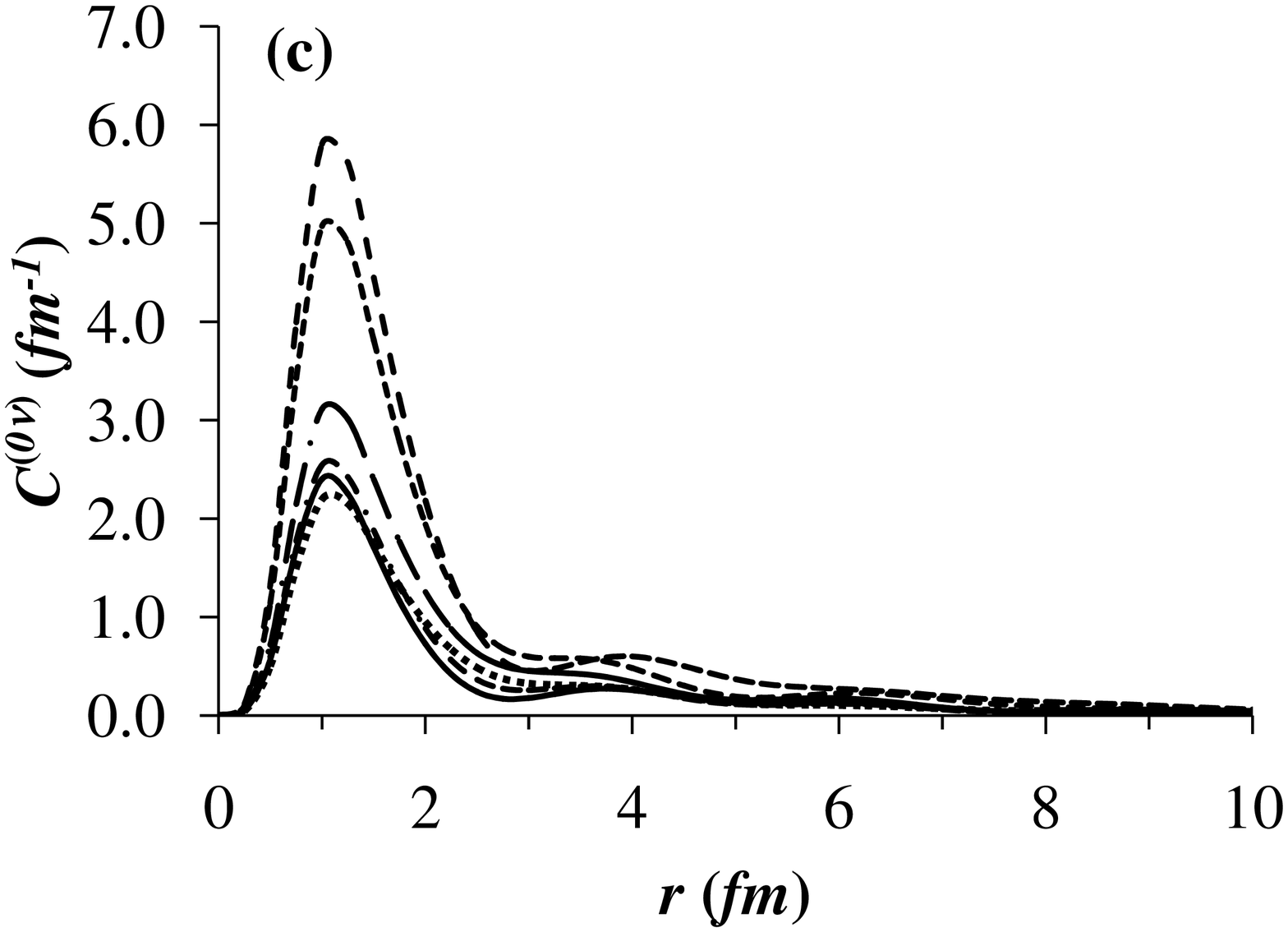} &
\includegraphics [scale=0.32]{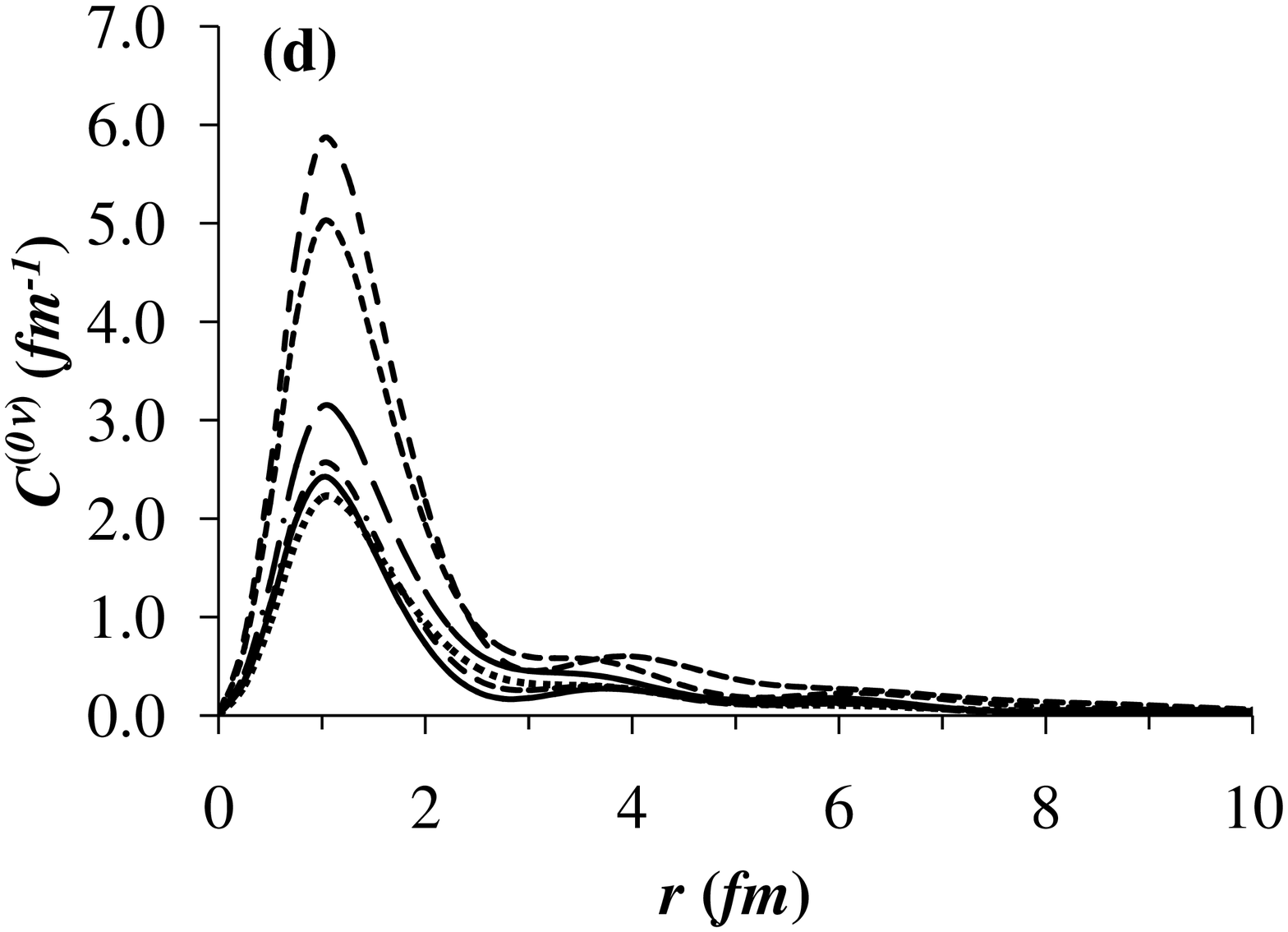} \\
\includegraphics [scale=0.32]{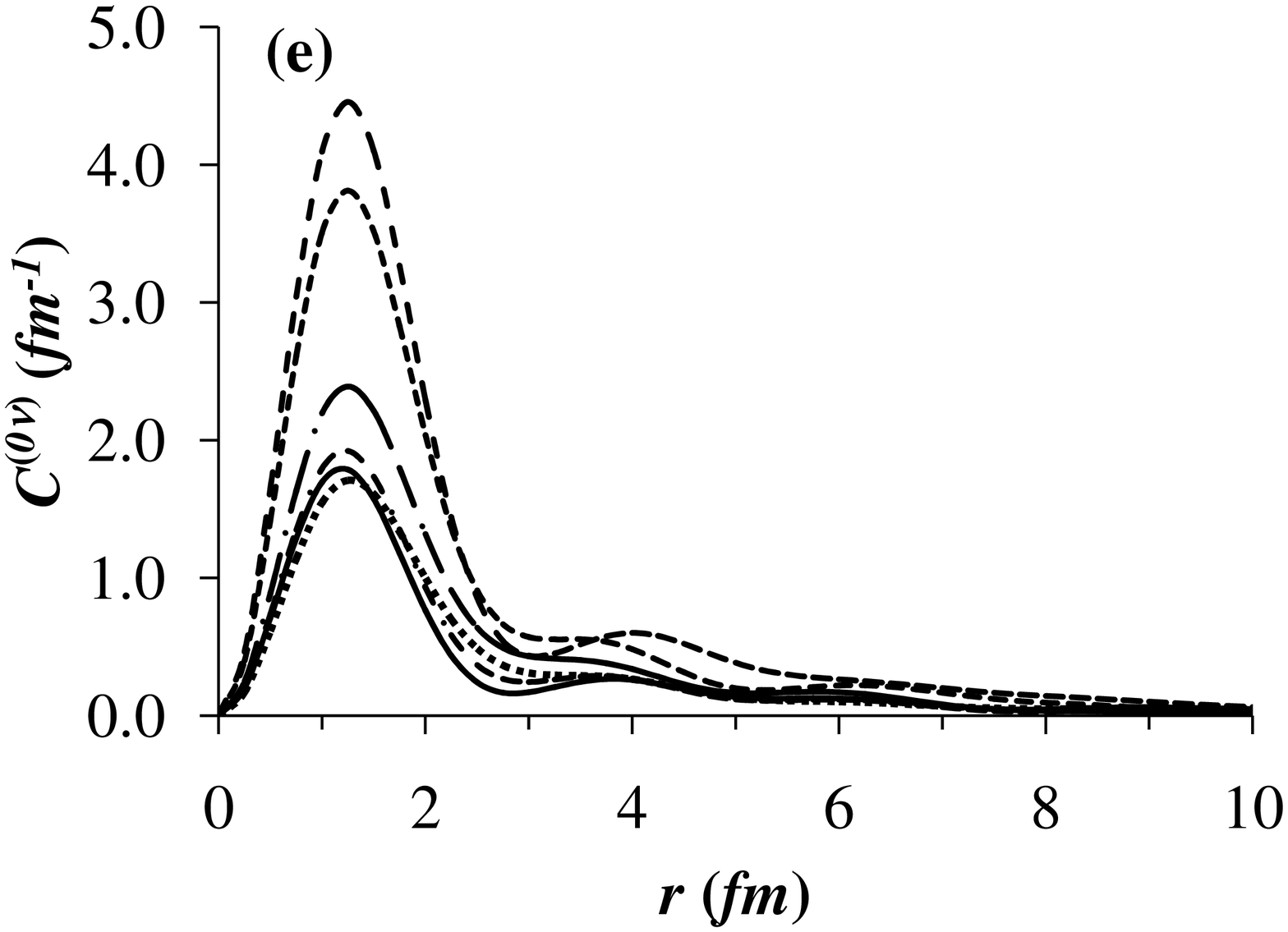} &
\includegraphics [scale=0.32]{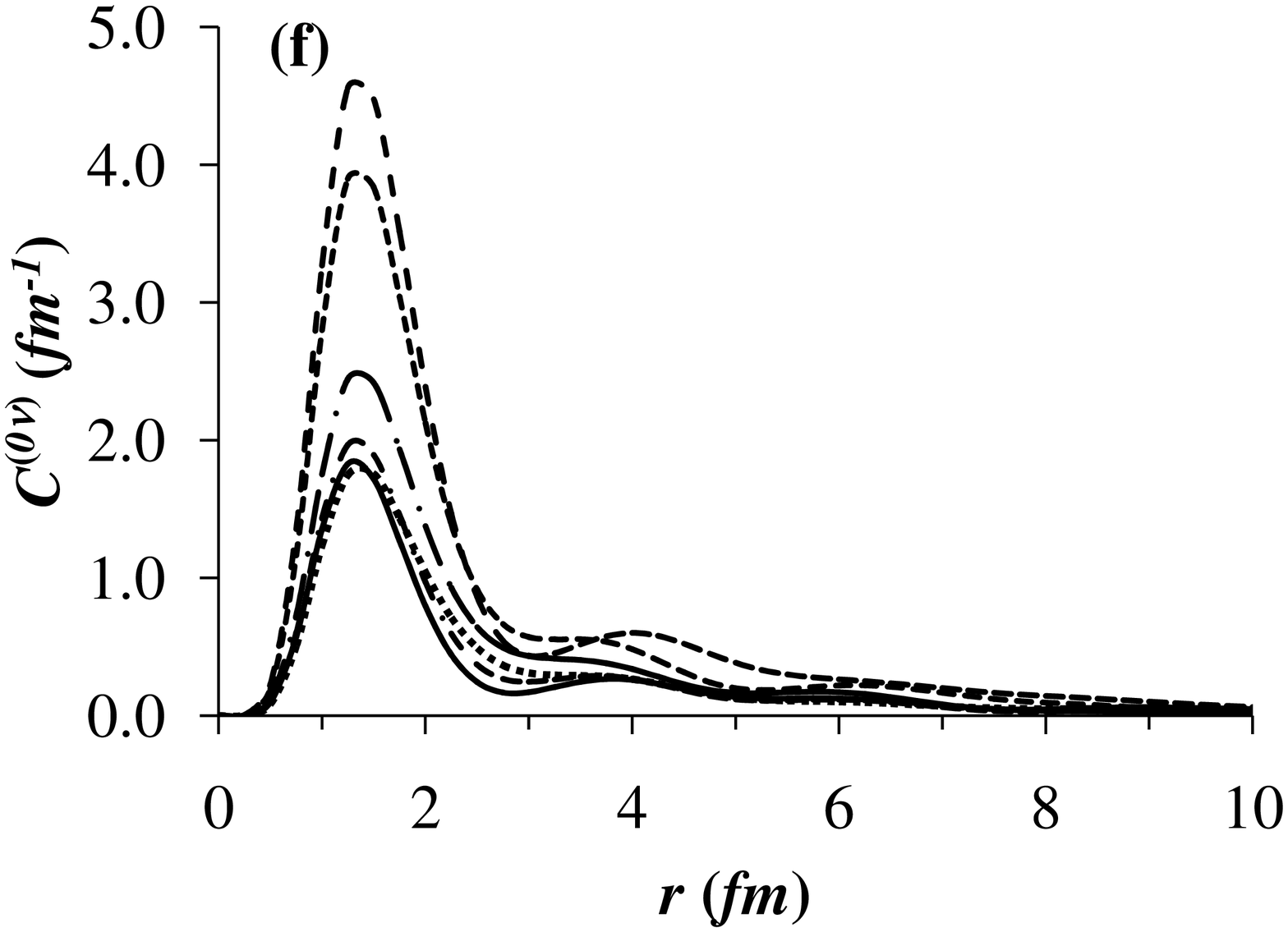} \\
\includegraphics [scale=0.32]{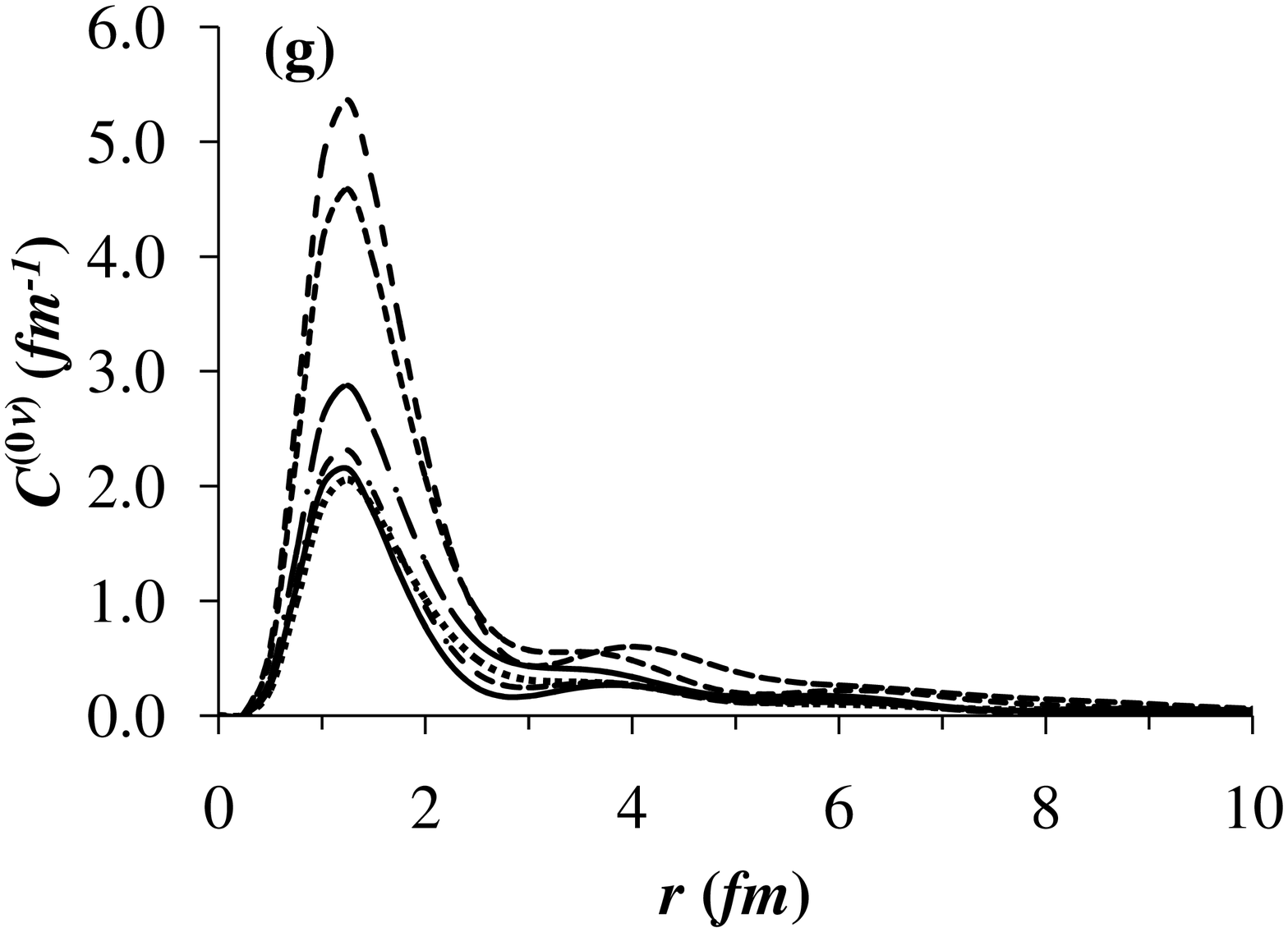} &
\includegraphics [scale=0.32]{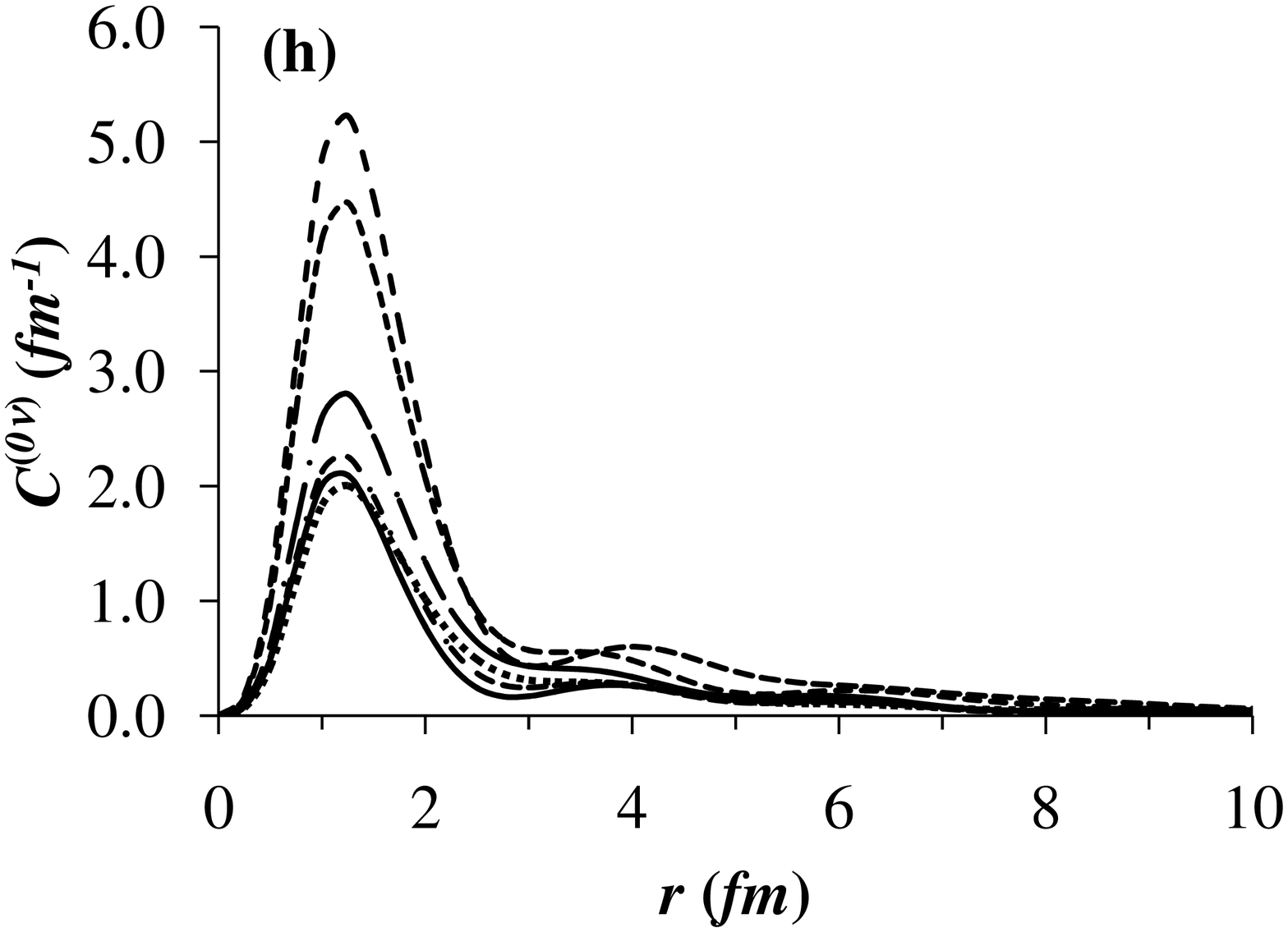} \\
\end{tabular}
\caption{Radial dependence of $C^{(0\nu)}(r)$ for the
$\left( \beta^{-}\beta ^{-}\right) _{0\nu }$ decay of $^{96}$Zr,
$^{100}$Mo, $^{110}$Pd, $^{128,130}$Te and $^{150}$Nd isotopes.
In this Fig., (a), (b), (c) and (d) correspond to P, P+SRC1, P+SRC2 and P+SRC3,
respectively. Further, (e), (f), (g) and (h) are for F, F+SRC1, F+SRC2 and 
F+SRC3, respectively.}
\label{fig2}
\end{figure*}
\begin{figure}[htbp]
\begin{tabular}{c}
\includegraphics [scale=0.38]{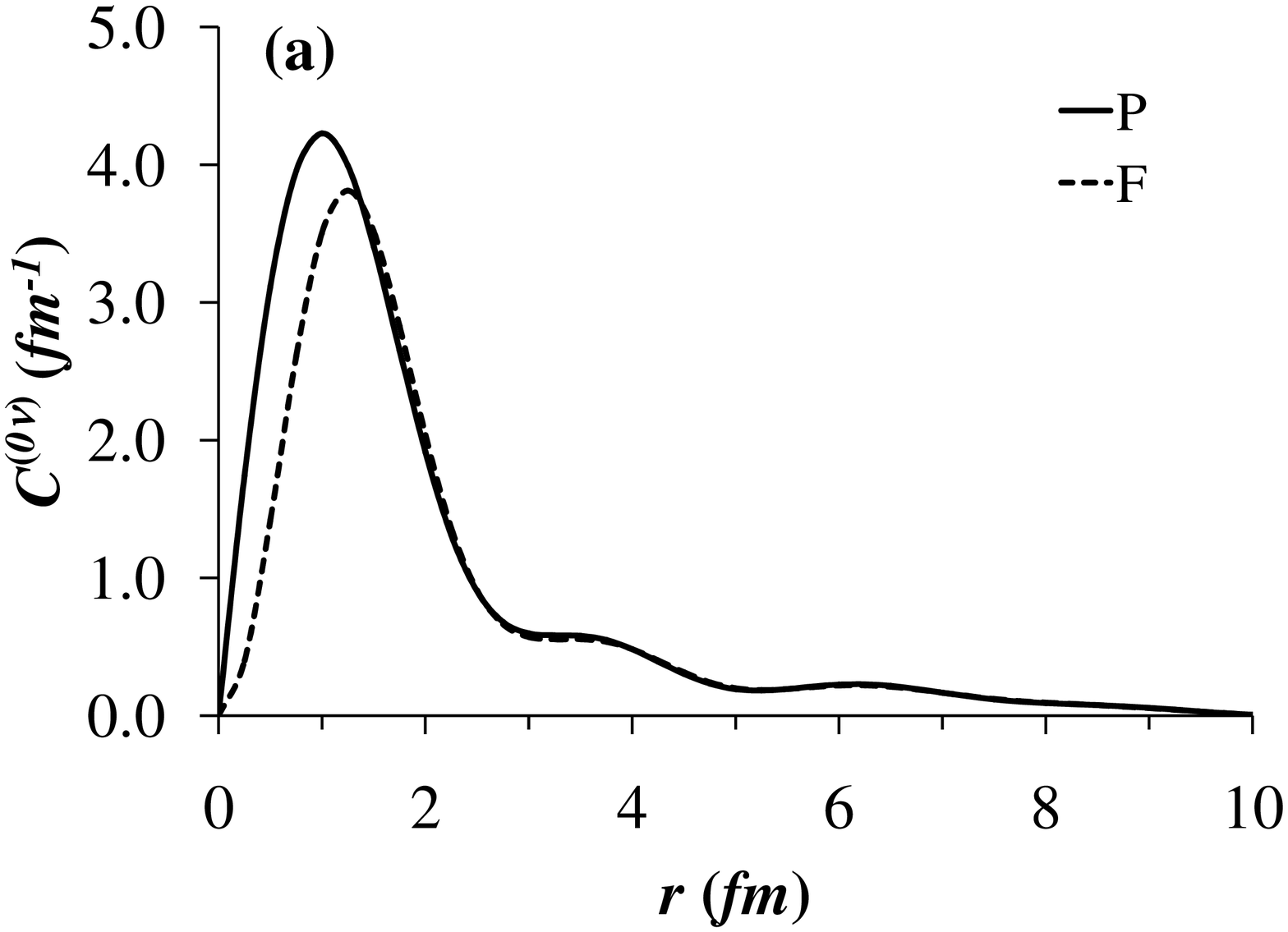} \\
\includegraphics [scale=0.38]{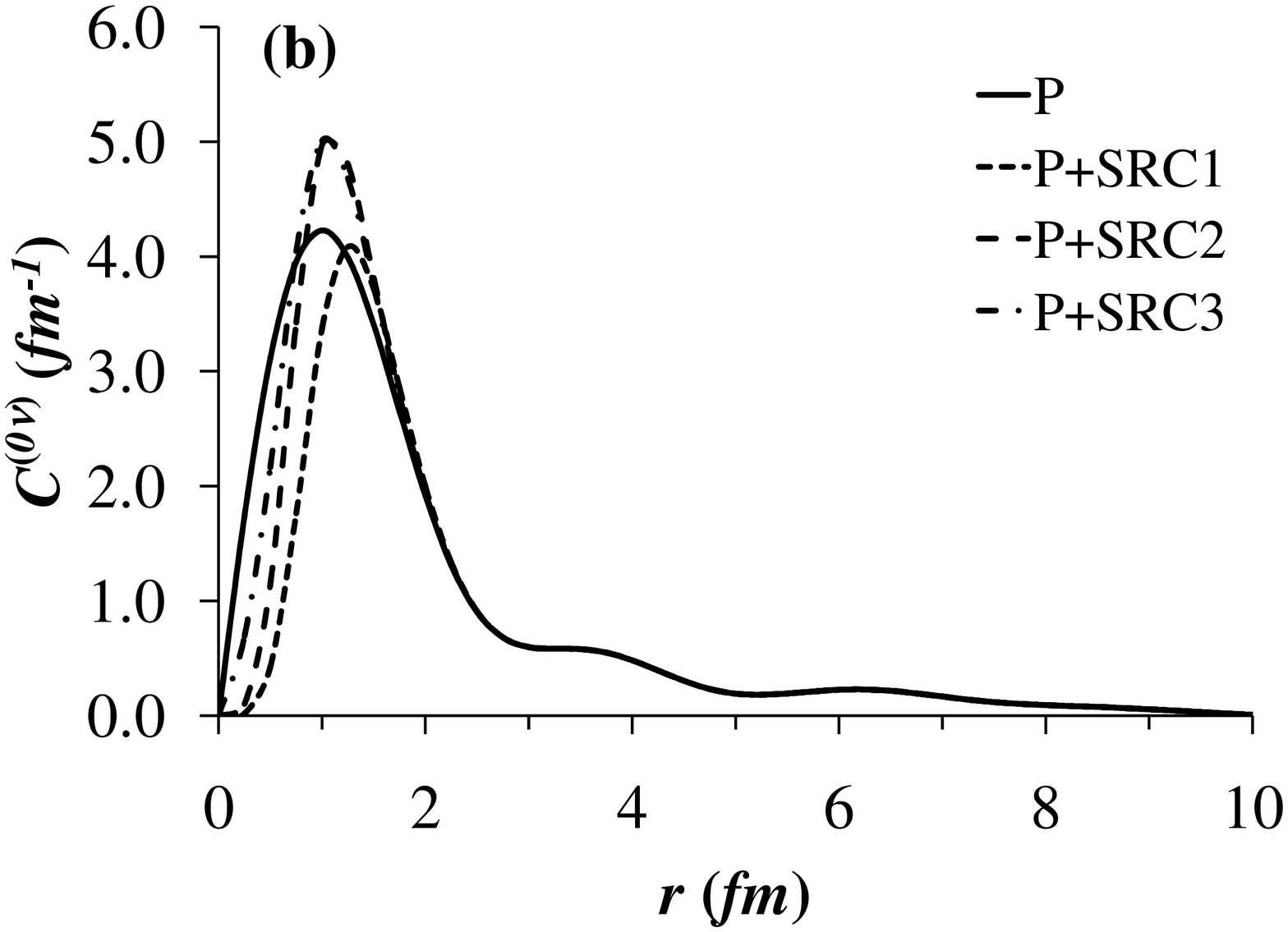} \\
\includegraphics [scale=0.38]{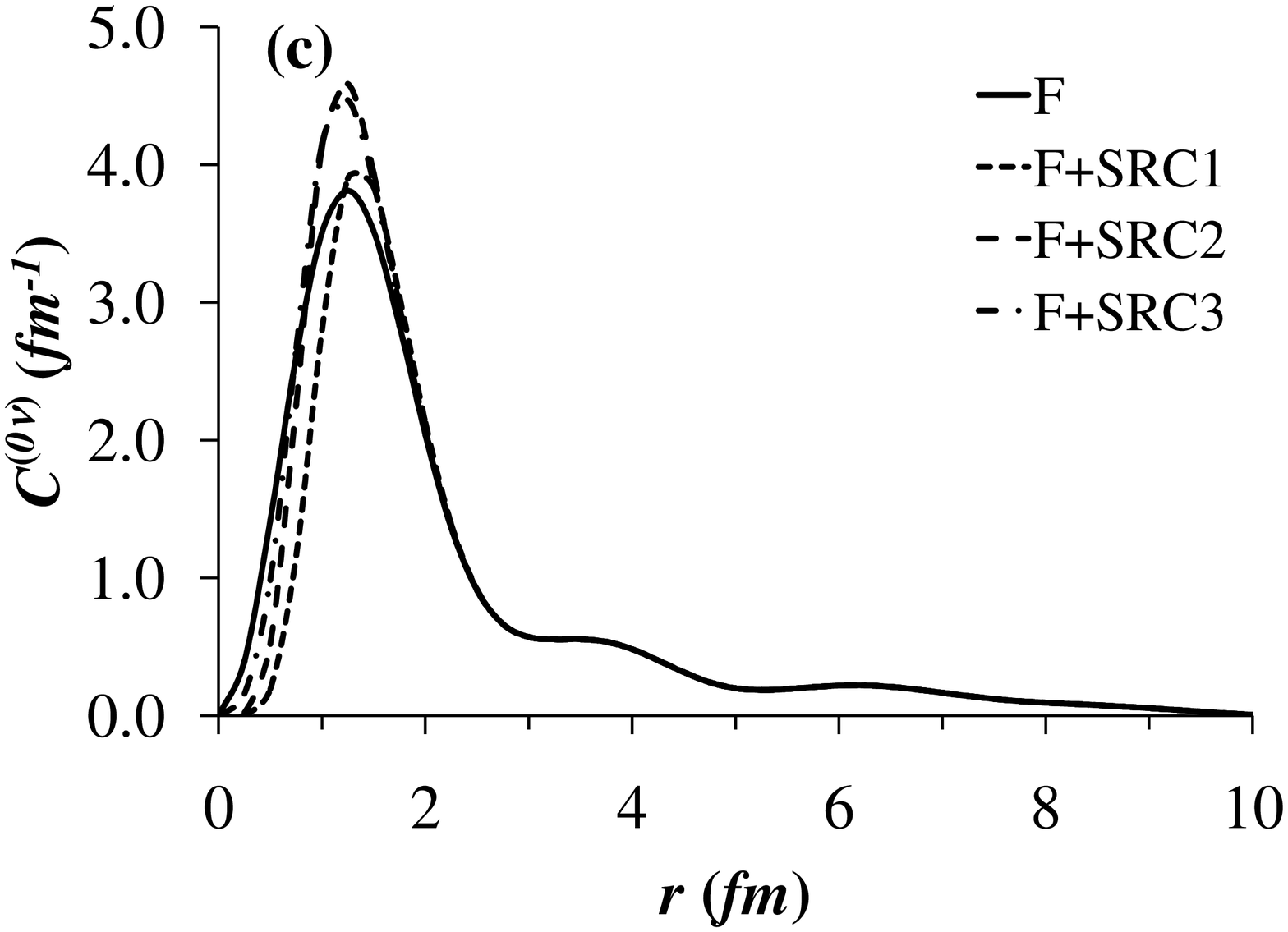} \\
\end{tabular}
\caption{Radial dependence of $C^{(0\nu)}(r)$ for the
$\left( \beta^{-}\beta ^{-}\right) _{0\nu }$ decay of $^{100}$Mo isotope.}
\label{fig3}
\end{figure}

The radial evolution of $M^{\left( 0\nu \right) }$ has been studied in the
QRPA by \v{S}imkovic $et$ $al.$ \cite{simk08} and in the ISM by Men\'{e}ndez 
$et$ $al.$ \cite{mene09} by defining 
\begin{equation}
M^{\left( 0\nu \right) }=\int C^{\left( 0\nu \right) }\left( r\right) dr.
\end{equation}
In both QRPA and ISM calculations, it has been established that
the contributions of decaying pairs coupled to $J=0$ and $J>0$ almost cancel
beyond $r\approx 3$ fm and the magnitude of $C^{\left( 0\nu \right) }$
for all nuclei undergoing $\left( \beta ^{-}\beta ^{-}\right) _{0\nu }$
decay are the maximum about the internucleon distance $r\approx 1$ fm.
In Fig.~\ref{fig2}, we plot the radial dependence of $C^{\left( 0\nu \right) }$ due to
$PQQ1$ parametrization of the effective two body interaction for
six nuclei, namely $^{96}$Zr, $^{100}$Mo, $^{110}$Pd, $^{128,130}$Te and $%
^{150}$Nd. 
The radial evolution of $M^{\left( 0\nu \right) }$
is studied for eight cases, namely P, P+SRC1, P+SRC2, P+SRC3, F,
F+SRC1, F+SRC2 and F+SRC3. In addition, the effects due to the finite size 
and SRC are made more transparent in Fig.~\ref{fig3} by plotting them for different 
combinations of P, F and SRC.   
In case of point nucleons, it is noticed that the $C^{\left( 0\nu
\right) }$ are peaked at $r=1.0$ fm and with the addition of
SRC1, the peak shifts to 1.25 fm. However, the magnitude of $C^{\left( 0\nu
\right) }$ are increased for SRC2 and SRC3 with unchanging peak position. In
the case of FNS, the $C^{\left( 0\nu \right) }$ are peaked at $r=1.25$ fm, 
which remains unchanged with the inclusion of SRC1, SRC2 and SRC3.
However, the magnitudes of $C^{\left( 0\nu \right) }$ change in the latter
three cases. The above observations also remain valid with the other three 
parametrizations of the effective two-body interaction. 

\subsection{Uncertainties in NTMEs}

To estimate the uncertainties associated with the NTMEs $M^{(0\nu )}$ 
for $\left( \beta ^{-}\beta ^{-}\right) _{0\nu }$ decay
calculated using the PHFB model, we evaluate 
their mean and the standard deviation,  defined as 
\begin{equation}
\overline{M}^{(0\nu )}=\frac{\sum_{i=1}^{N}M_{i}^{(0\nu )}}{N}
\end{equation}
and 
\begin{equation}
\Delta \overline{M}^{(0\nu )}=\frac{1}{\sqrt{N-1}}\left[
\sum_{i=1}^{N}\left( \overline{M}^{(0\nu )}-M_{i}^{(0\nu )}\right)
^{2}\right] ^{1/2}.
\end{equation}
Recently, it has been shown by \v{S}imkovic \textit{et al.} \cite{simk09} that the 
phenomenological Jastrow correlations with Miller-Spenser parametrization is a major
source of uncertainty. Therefore, it is more appropriate to consider SRC2 or SRC3 due to 
the Argonne V18 and CD-Bonn NN potentials, respectively. Based on these observations, we perform
the statistical analysis of two cases. In case I, we calculate
the average and variance of twelve NTMEs listed in the last three columns (F+S) of 
Table \ref{tab1} with the bare and quenched values of axial vector coupling constant 
$g_{A}=1.254$ and $g_{A}=1.0$, respectively. The average and standard deviations of eight
NTMES $M^{(0\nu )}$ due to SRC2 and SRC3 are similarly calculated in the case II.
The average NTMEs $\overline{M}^{(0\nu )}$ and standard deviations 
$\Delta \overline{M}^{(0\nu )}$ are presented in Table~\ref{tab3}.
It is noticed that the exclusion of  Miller-Spenser parametrization reduces the 
uncertainty by about 55\% in $^{96}$Zr to 4\% in $^{150}$Nd isotope.
In Table~\ref{tab4}, we present the average NTMEs $\overline{M}^{(0\nu )}$ of case II 
along with the recently reported results in ISM by Caurier \textit{et al.} \cite{caur08}, QRPA as well as
RQRPA by \v{S}imkovic \textit{et al.} \cite{simk09} and IBM by Barea and Iachello \cite{bare09}.
In spite of the fact that different model space, two-body interactions and SRC have been used in 
these models, the spread in the NTMEs turns out to be about a factor of 2.5.    
Further, we extract upper limits on the effective mass of light
neutrinos $\left\langle m_{\nu }\right\rangle $ from the largest observed
limits on half-lives $T_{1/2}^{0\nu }$ of $\left( \beta ^{-}\beta
^{-}\right) _{0\nu }$ decay using the phase space factors of Boehm and Vogel \cite{boeh92}.
It is observed that the extracted limits on $%
\left\langle m_{\nu }\right\rangle $ for $^{100}$Mo and $^{130}$Te nuclei
are $0.48_{-0.03}^{+0.04}-0.69_{-0.05}^{+0.05}$ and 
$0.30_{-0.02}^{+0.03}-0.42_{-0.04}^{+0.04}$ eV, respectively. In the last
column of Table~\ref{tab4}, the predicted half-lives of $\left( \beta ^{-}\beta
^{-}\right) _{0\nu }$ decay of $^{94,96}$Zr, $^{98,100}$Mo, $^{110}$Pd, $%
^{128,130}$Te and $^{150}$Nd isotopes are given for $\left\langle m_{\nu
}\right\rangle =50$ meV.

\section{CONCLUSIONS}

We have studied the $\left( \beta ^{-}\beta ^{-}\right) _{0\nu }$ \ decay 
of $^{94,96}$Zr, $^{98,100}$Mo, $^{104}$Ru, $^{110}$Pd, $^{128,130}$Te and $^{150}$Nd
isotopes in the light Majorana neutrino mass mechanism 
using a set of PHFB wave functions.
The reliability of wave functions generated with $PQQ1$ and $PQQHH1$ interactions has 
been tested in previous works by
calculating the yrast spectra, reduced $B(E2$:$0^{+}\rightarrow 2^{+})$
transition probabilities, static quadrupole moments $Q(2^{+})$ and $g$%
-factors $g(2^{+})$ of participating nuclei in $\left( \beta ^{-}\beta
^{-}\right) _{2\nu }$ \ decay as well as $M_{2\nu }$ and comparing them with
the available experimental data \cite{chan05,sing07}. 
An overall agreement between the calculated and observed spectroscopic properties 
as well as $M_{2\nu }$ suggests that the PHFB wave functions generated by fixing 
$\chi _{pn}$ or $\chi _{pp}$ to reproduce the $E_{2^{+}}$ are reasonably reliable.

In the present work,
 NTMEs $M^{\left( 0\nu \right) }$  
were calculated employing the PHFB model with four different
parameterizations of the pairing plus multipolar type of effective two body
interaction and two(three) different parameterizations of the short range correlations. 
It was found that the NTMEs $M^{\left( 0\nu \right) }$ change by about 4--14(10--15)\%.

The mean and standard deviations were evaluated for the NTMEs $M^{\left( 0\nu \right) }$
calculated with dipole form factor and with and without Miller-Spencer parametrization of 
short range correlations,  
which were employed to estimate the $\left( \beta ^{-}\beta ^{-}\right) _{0\nu }$ 
decay half-lives $T_{1/2}^{0\nu }$ for both $g_{A}=1.254$ and $g_{A}=1.0$. 
The largest standard deviation, interpreted as theoretical uncertainty, 
turns out to be of the order of 15\% in the case of $^{150}$Nd isotope. 
We have also extracted limits on the
effective mass of light Majorana neutrinos $\left\langle m_{\nu }\right\rangle $ from the
available limits on experimental half-lives $T_{1/2}^{0\nu }$ using average
NTMEs $\overline{M}^{(0\nu )}$ calculated in the PHFB model. In the case of $%
^{130}$Te isotope, one obtains the best limit on the effective neutrino mass 
$\left\langle m_{\nu }\right\rangle
<0.30_{-0.02}^{+0.03}-0.42_{-0.04}^{+0.04}$ eV from the observed limit on the
half-lives $T_{1/2}^{0\nu }>3.0\times 10^{24}$ yr of $\left( \beta ^{-}\beta
^{-}\right) _{0\nu }$ decay \cite{arna08}.

\noindent{\textbf{Note:}\textit{ Due to an error in one equation, the NTMEs 
$M_F$, $M_{GT}$,  $M^{(0\nu)}$, $M_{Fh}$, $M_{GTh}$ and $M_{N}^{(0\nu)}$ given 
in Ref. \cite{chat08} must be multiplied by a factor of 2. It implies that the 
limits on the effective light neutrino
mass $<m_{\nu}>$ must be reduced by a factor of 2 whereas the limits on effective heavy 
neutrino mass $<M_N>$ must be multiplied by a factor of 2. In both cases the limits 
are twice more stringent.}}

\begin{acknowledgments}
This work is partially supported by DST, India vide sanction No.
SR/S2/HEP-13/2006, Conacyt-M\'{e}xico, FONCICYT project 94142 and DGAPA-UNAM.
\end{acknowledgments}

\end{document}